\documentclass{aa}
\usepackage{graphicx}
\usepackage{txfonts}
\begin{document}
\title{
High spectral resolution imaging of the dynamical atmosphere of 
the red supergiant Antares in the CO first overtone lines with VLTI/AMBER
\thanks{
Based on AMBER observations made with the Very Large Telescope 
Interferometer of the European Southern Observatory. 
Program ID: 083.D-0333(A/B) (AMBER Guaranteed Time Observation), 085.D-0085(A/B)}
}

\author{K.~Ohnaka\inst{1} 
\and
K.-H.~Hofmann\inst{1} 
\and
D.~Schertl\inst{1}
\and
G.~Weigelt\inst{1} 
\and
C.~Baffa\inst{2}
\and
A.~Chelli\inst{3}
\and
R.~Petrov\inst{4}
\and
S.~Robbe-Dubois\inst{4}
}

\offprints{K.~Ohnaka}

\institute{
Max-Planck-Institut f\"{u}r Radioastronomie, 
Auf dem H\"{u}gel 69, 53121 Bonn, Germany\\
\email{kohnaka@mpifr.de}
\and
INAF-Osservatorio Astrofisico di Arcetri, Instituto Nazionale di 
Astrofisica, Largo E. Fermi 5, 50125 Firenze, Italy
\and
UJF-Grenoble 1 / CNRS-INSU, Institut de Plan\'etologie et d'Astrophysique de
Grenoble (IPAG) UMR 5274, Grenoble, F-38041, France
\and
Observatoire de la Cote d'Azur, Departement FIZEAU,
Boulevard de l'Observatoire, B.P. 4229, 06304 Nice Cedex 4,
France
}

\date{Received / Accepted }

\abstract
{}
{
We present aperture-synthesis 
imaging of the red supergiant Antares ($\alpha$~Sco) in the CO first 
overtone lines.  
Our goal is to probe the structure and dynamics of the outer atmosphere. 
}
{
Antares was observed between 2.28~\mbox{$\mu$m}\ and 2.31~\mbox{$\mu$m}\ with VLTI/AMBER 
with spectral resolutions of up to 12000 and 
angular resolutions as high as 7.2~mas at two epochs with a time interval of 
one year.  
}
{
The reconstructed images in individual CO lines reveal that the 
star appears differently in the blue wing, line center, and red wing.  
In 2009, the images in the line center and red wing show an asymmetrically 
extended component, while the image in the blue wing 
shows little trace of it.  
In 2010, however, the extended component appears in the line center and 
blue wing, 
and the image in the red wing shows only a weak signature of the extended 
component.  
Our modeling of these AMBER data suggests that there is an outer 
atmosphere (MOLsphere) 
extending to 1.2--1.4~\mbox{$R_{\star}$}\ with CO column densities of 
$(0.5$--$1)\times10^{20}$~\mbox{cm$^{-2}$}\ and a temperature of 
$\sim$2000~K. 
The CO line images observed in 2009 can be explained by a model in which 
a large patch or clump of CO gas is infalling at only 0--5~\mbox{km s$^{-1}$}, while 
the CO gas in the remaining region is moving outward much faster at 
20--30~\mbox{km s$^{-1}$}.  
The images observed in 2010 suggest that a large clump of 
CO gas is moving outward at 0--5~\mbox{km s$^{-1}$}, while the CO gas in the 
remaining region is infalling much faster at 20--30~\mbox{km s$^{-1}$}.  
In contrast to the images in the CO lines, 
the AMBER data in the {{\em continuum}} show only 
a slight deviation from limb-darkened disks and 
only marginal time variations.  
We derive a limb-darkened disk diameter of $37.38 \pm 0.06$~mas and 
a power-law-type limb-darkening parameter of $(8.7 \pm 1.6) \times 
10^{-2}$ (2009) and $37.31 \pm 0.09$~mas and 
$(1.5 \pm 0.2) \times 10^{-1}$ (2010).  
{{We also obtain an effective temperature of $3660\pm 120$~K 
(the error includes the effects of the temporal flux variation that 
is assumed to be the same as Betelgeuse) 
and a luminosity of $\log L_{\star}/\mbox{$L_{\sun}$}=4.88 \pm 0.23$.  Comparison with 
theoretical evolutionary tracks suggests a mass of $15\pm5$~\mbox{$M_{\sun}$}\ 
with an age of 11--15~Myr, which is consistent with 
the recently estimated age for the Upper Scorpius OB association. }}
}
{
The properties of the outer atmosphere of Antares are similar to 
those of another well-studied red supergiant, Betelgeuse. 
The density of the extended outer atmosphere of Antares and Betelgeuse is 
higher than predicted by the current 3-D convection 
simulations by at least six orders of magnitude, 
implying that convection alone cannot explain the formation of the 
extended outer atmosphere.  
}

\keywords{
infrared: stars --
techniques: interferometric -- 
stars: supergiants  -- 
stars: late-type -- 
stars: atmospheres -- 
stars: individual: Antares
}   

\titlerunning{High spectral resolution imaging of the dynamical 
atmosphere of Antares in the CO first overtone lines}
\authorrunning{Ohnaka et al.}
\maketitle

\section{Introduction}
\label{sect_intro}

The mass loss in the red supergiant (RSG) stage significantly affects 
the final fate of massive stars.  
For example, the mass loss in the RSG stage seems to be a key to 
understanding the progenitors of the most common core-collapse supernovae 
(SNe) Type IIp (e.g., Smartt et al. \cite{smartt09}).  
However, the mass loss in RSGs is still poorly understood.  
As Harper (\cite{harper10}) stresses, there are currently 
no satisfactory theories for the RSG mass loss.  Even the driving force 
of the stellar winds has not yet been identified.  

For understanding the wind acceleration mechanism, it is important to 
study the region between the photosphere and the innermost circumstellar 
envelope, where the stellar winds are expected to be accelerated.  
Various observations suggest the complicated nature of this region. 
Tuthill et al. (\cite{tuthill97}) have discovered asymmetric structures 
on the surface of three 
well-studied RSGs, \object{Betelgeuse} ($\alpha$~Ori), \object{Antares} 
($\alpha$~Sco), and \object{$\alpha$~Her}.  
They interpret these asymmetric structures as the presence of one to 
three hot spots.  
The recent near-IR interferometric observations of Betelgeuse 
by Haubois et al. (\cite{haubois09}) also show surface 
inhomogeneities.  
The detection of \mbox{H$_2$O}\ in Betelgeuse and Antares at 12~\mbox{$\mu$m}\ 
(Jennings \& Sada \cite{jennings98}), as well as in the near-IR and 
in the 6~\mbox{$\mu$m}\ region (Tsuji \cite{tsuji00a}; \cite{tsuji00b}), 
suggests that there is an outer atmosphere extending 
to 1.3--2 stellar radii with temperatures of 1500--2000~K, the so-called 
MOLsphere as coined by Tsuji (\cite{tsuji00b}).

The radio observations of Betelgeuse and Antares 
also show neutral gas extending 
to several stellar radii (Lim et al. \cite{lim98}; Harper et al. 
\cite{harper10}).  
On the other hand, spatially resolved observations of RSGs 
in the UV, as well as in the H$\alpha$ line, suggest that the chromosphere 
also extends to several stellar radii 
(Gilliland \& Dupree \cite{gilliland96}; White et al. \cite{white82}; 
Hebden et al. \cite{hebden87}).  
Harper et al. (\cite{harper01}) propose that the chromosphere has a 
small filling factor embedded in the neutral/molecular gas.  
However, the physical mechanism responsible for this inhomogeneous, 
multicomponent outer atmosphere is by no means clear.  
Inhomogeneous structures are also found farther out from the star. 
The near- and mid-IR imaging of Betelgeuse 
(Kervella et al. \cite{kervella09}, \cite{kervella11}) and 
Antares (Bloemhof et al. \cite{bloemhof84}; \cite{bloemhof95}; 
Marsh et al. \cite{marsh01}) reveals a clumpy structure of the 
circumstellar envelope.  

For a better understanding of the mass-loss mechanism in RSGs, 
it is essential to probe the dynamics of the inhomogeneous, multicomponent 
outer atmosphere. 
We carried out high spatial and high spectral resolution observations 
of Betelgeuse in the CO first overtone lines near 2.3~\mbox{$\mu$m}\ using 
VLTI/AMBER (Ohnaka et al. \cite{ohnaka09}; \cite{ohnaka11}, hereafter Paper I 
and II, respectively).  
With the stellar disk spatially resolved with an angular resolution of 
9.8~mas (the highest resolution ever achieved at any wavelength for this 
well-studied star), AMBER's high spectral resolution of up to 12000 
enabled us to spatially resolve, for the first time, the gas motions over 
the surface of a star other than the Sun.  These AMBER observations have 
revealed that the star appears differently in the blue and red wings of 
the individual CO lines due to an inhomogeneous velocity field in the 
photosphere and MOLsphere.   

In this paper, 
we present high spectral and high spatial resolution VLTI/AMBER observations 
of another well-studied RSG, Antares (M1.5Iab-b), in the CO first overtone 
lines.  
Antares is similar to Betelgeuse (M2Iab:) in terms of the spectral 
type, although it is slightly less luminous and less massive than 
Betelgeuse (see Sect.~\ref{sect_res_param}). 
The mass-loss rate of Antares is estimated to be 
$\sim \!\! 2 \times 10^{-6}$~\mbox{$M_{\sun}$}PERYR\ (Braun et al. \cite{braun12}), 
which is comparable to that of Betelgeuse 
($\sim \!\! 3 \times 10^{-6}$~\mbox{$M_{\sun}$}PERYR, Harper et al. \cite{harper01}).  
However, the IR excess and the 10~\mbox{$\mu$m}\ silicate emission of 
Antares are much weaker than those of Betelgeuse (e.g., Monnier et al. 
\cite{monnier98}; Verhoelst et al. \cite{verhoelst09}).  
Our goal is to spatially resolve the gas motions in the outer atmosphere 
in this second RSG and to examine whether the findings for Betelgeuse 
are common among RSGs.  
The paper is organized as follows.  
The observations and data reduction as well as image reconstruction 
from the interferometric data are described in Sect.~\ref{sect_obs}.  
We present the results in Sect.~\ref{sect_res} followed by 
the modeling of the data in Sect.~\ref{sect_modeling} and the interpretation 
of the results in Sect.~\ref{sect_discuss}.  Conclusions are given in 
Sect.~\ref{sect_concl}.

\section{Observations}
\label{sect_obs}

\begin{figure*}
\sidecaption
\rotatebox{0}{\includegraphics[width=12cm]{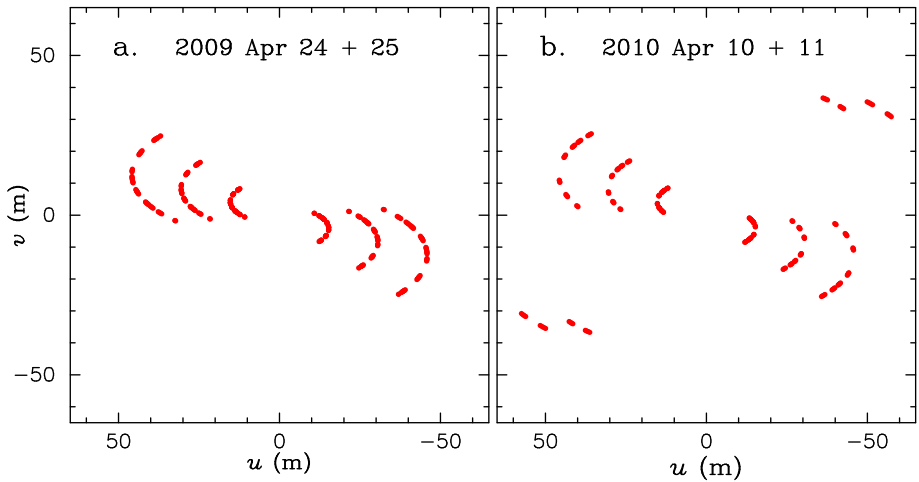}}
\caption{
The $u\varv$ coverage of our AMBER observations of Antares in 2009 ({\bf a}) and 
2010 ({\bf b}).  
}
\label{uv_coverage}
\end{figure*}

\subsection{AMBER observations and data reduction}
AMBER (Petrov et al. \cite{petrov07}) is a VLT spectro-interferometric 
instrument that operates in the near-IR (1.3---2.4~\mbox{$\mu$m}) and combines 
three 8.2~m Unit Telescopes (UTs) or 1.8~m Auxiliary Telescopes (ATs). 
AMBER records spectrally-dispersed interferograms 
with a spectral resolution of up to 12000. 
AMBER measures the amplitude of the Fourier transform (so-called visibility 
or visibility amplitude), closure phase (CP) and differential phase (DP), 
with the usual (i.e., spatially unresolved) spectrum.  
The CP is the sum of the measured Fourier phases around a closed triangle of 
baselines (i.e., $\varphi_{12} + \varphi_{23} + \varphi_{31}$), which is not 
affected by the atmospheric turbulence.  Therefore, it is essential for 
aperture-synthesis imaging in optical/IR interferometry. 
The CP is zero or $\pi$ for point-symmetric objects and 
non-zero and non-$\pi$ CPs indicate an asymmetry of the object. 
The DP provides information about the wavelength dependence of the 
photocenter shift of the object within spectral features.  

Antares was observed on 2009 April 24 and 25 and, one year later, on 
2010 April 10 and 11, with AMBER using the E0-G0-H0 (16--32--48~m) array 
configuration and the E0-G0-I1 (16--57--69~m) configuration 
(Program IDs: 083.D-0333, 085.D-085, P.I.: K.~Ohnaka). 
As in Papers I and II, 
we used the high resolution mode in the $K$-band (HR\_K) with a spectral 
resolution of 12000 and covered wavelengths from 2.28 to 2.31~\mbox{$\mu$m}\ to 
observe the CO first overtone lines near the (2,0) band head.  
We detected fringes 
on all three baselines without using the VLTI fringe tracker FINITO 
(Antares saturates FINITO) thanks to the extremely high brightness of 
Antares ($K = -4.1$).  
In 2009 and 2010, we observed on two half nights each and obtained 
65 and 44 data sets, respectively.  
The data sets taken more than $\sim$2 minutes apart were treated as 
separate data sets, because Antares is 
strongly resolved, and the visibility varies noticeably even with a slight 
change in the projected baseline length (see Fig.~\ref{vis_continuum}).  
Figure~\ref{uv_coverage} shows the $u\varv$ coverage of our observations 
in 2009 and 2010.  
All data sets were taken with a Detector Integration Time (DIT) of 130~ms.  
A summary of the observations is given in Tables~\ref{obslog2009} and 
\ref{obslog2010}.

For the reduction of the AMBER data, we used amdlib ver. 2.2\footnote{
Available at http://www.jmmc.fr/data\_processing\_amber.htm}, 
which is based on the P2VM algorithm (Tatulli et al. \cite{tatulli07}).  
In order to improve the signal-to-noise ratios (S/N), we applied 
a binning in the spectral direction to all the raw data (object, dark, 
sky, and P2VM calibration data) using a running box car function as 
described in Paper I.  
The binning down to a spectral resolution of 8000 provided 
an S/N sufficient for the image reconstruction. 
One of the parameters in the reduction with amdlib is the frame selection 
criterion.  
In each data set, 
we checked for a systematic difference in the calibrated results 
by taking the best 20\%, 50\%, and 80\% of all frames in terms of the fringe 
S/N (Tatulli et al. \cite{tatulli07}). 
The difference between the results obtained with the best 20\% and 80\% 
of frames is negligible, and therefore, we took the best 80\% of all frames 
for our final results.  
The errors of the resulting visibilities, DPs, and CPs were estimated 
in the manner described in Paper I.

\object{$\alpha$~Cen A} (G2V) and \object{$\alpha$~Cen B} 
(K1V) were observed for the calibration of the interferometric data of 
Antares. 
$\alpha$~Cen is a triple system consisting of $\alpha$ Cen~A, 
$\alpha$~Cen B, and Proxima Cen.  However, we always had only one 
component (A or B) in the field of view of AMBER.  
We adopted angular diameters of $8.314 \pm 0.016$ and $5.856\pm0.027$~mas 
derived by Kervella et al. (\cite{kervella03}) for $\alpha$~Cen A and B, 
respectively.  
We also used $\alpha$~Cen A as a spectroscopic standard star to remove 
telluric lines from the observed spectra of Antares.  
However, the spectrum of $\alpha$~Cen A shows weak CO first overtone 
lines.  In this case, as described in Ohnaka et al. (\cite{ohnaka12}), 
the calibrated spectrum of the science target is derived as 
$F_{\star}^{\rm sci} = F_{\rm obs}^{\rm sci} / (F_{\rm obs}^{\rm
  cal}/F_{\star}^{\rm cal})$, 
where $F_{\star}^{\rm sci(cal)}$ and $F_{\rm obs}^{\rm sci(cal)}$ denote 
the true and observed (i.e., including the atmospheric transmission 
and the detector's response) spectra of the science target 
(or calibrator), respectively.  
We used the high resolution solar spectrum presented by Wallace \& 
Hinkle (\cite{wallace96}) to estimate the true spectrum of the 
calibrator $\alpha$~Cen A, because this star has the same spectral 
type as the Sun.  

\begin{figure*}
\sidecaption
\rotatebox{0}{\includegraphics[width=12cm]{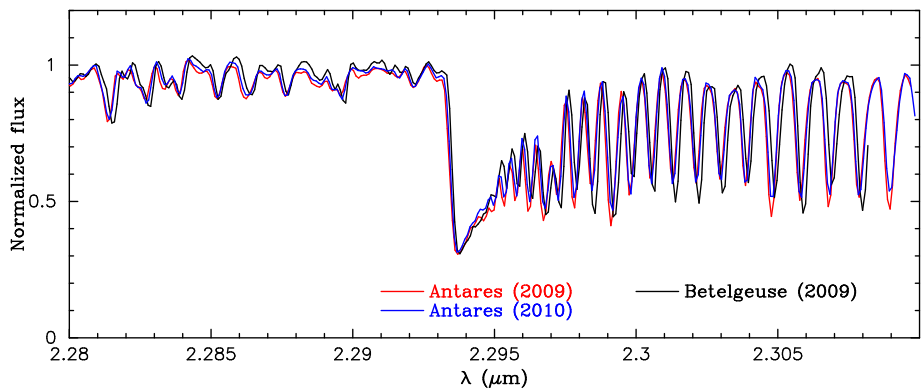}}
\caption{
Spectra of Antares observed in 2009 and 2010, together with the spectrum 
of Betelgeuse from Paper II (the one obtained in 2009). 
}
\label{specplot}
\end{figure*}

\begin{figure*}
\resizebox{\hsize}{!}{\rotatebox{-90}{\includegraphics{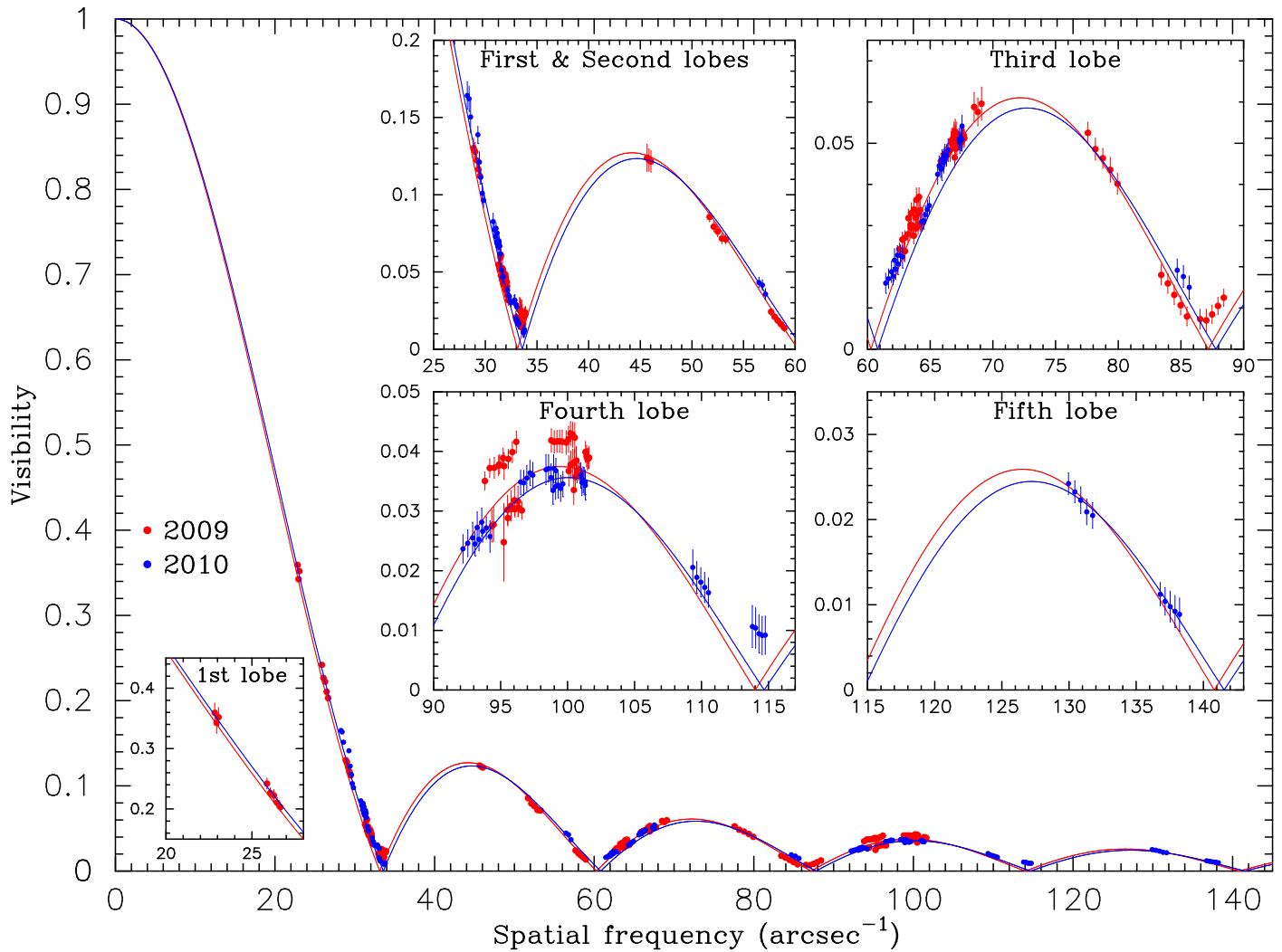}}}
\caption{
The $K$-band continuum visibilities of Antares observed in 2009 (red dots) 
and 2010 (blue dots).  The fit with the power-law-type limb-darkened disk 
for the 2009 and 2010 data is shown by the red and blue solid lines, 
respectively. 
The limb-darkened disk has angular diameters of 37.38 and 37.31~mas and 
limb-darkening parameters of 0.087 and 0.15 for the 2009 and 2010 data, 
respectively.  
The insets show the enlarged views of the individual visibility lobes. 
}
\label{vis_continuum}
\end{figure*}

The wavelength calibration was done by using 
the telluric lines in the spectrum of $\alpha$~Cen A.  
As a template of the telluric lines, 
we convolved the atmospheric transmission spectra measured at the Kitt Peak 
National Observatory\footnote{http://www.eso.org/sci/facilities/paranal/instruments/isaac/tools/\\spectra/atmos\_S\_K.fits} 
to match the spectral resolutions of the data.  
The uncertainty in wavelength calibration is 
$1.7 \times 10^{-5}$~\mbox{$\mu$m}\ ($\sim$2~\mbox{km s$^{-1}$}).  
The wavelength scale was converted to the heliocentric frame and 
then to the laboratory frame using the heliocentric velocity of 
$-3.5$~\mbox{km s$^{-1}$}\ measured for Antares (Gontscharov \cite{gontscharov06}).  
Figure~\ref{specplot} shows the spectra of Antares observed in 2009 
and 2010, together with the spectrum of Betelgeuse obtained in 2009 
(Paper II).  These spectra are binned to the same spectral resolution of 
8000.  
The observed spectra of Antares are similar to that of Betelgeuse 
and show little time variation within an interval of one year.  
However, despite this absence of time variation in the observed spectra, 
our spectro-interferometric observations have detected significant 
time variation in the velocity field of the atmosphere, as presented in 
Sect.~\ref{subsect_imaging}.  

\subsection{Image reconstruction}

We used the MiRA package ver.0.9.9\footnote{http://www-obs.univ-lyon1.fr/labo/perso/eric.thiebaut/mira.html}(Thi\'ebaut 
\cite{thiebaut08}) to reconstruct an image 
from the interferometric measurements at each spectral channel.  
As described in Paper II, we first carried out the image reconstruction 
using simulated data. 
As described in Appendix \ref{appendix_simtests}, 
we generated simulated stellar images (e.g., 
limb-darkened disk or stellar disk with inhomogeneities and/or an extended 
component) and 
computed visibilities and CPs with realistic noise at the same $u\varv$ 
points as in our AMBER observations of Antares.  
With the true \mbox{image} known, 
we searched for the appropriate reconstruction 
parameters, such as the initial model and regularization scheme.  
We also applied the self-calibration technique described in Paper II, 
which restores the Fourier phase from the differential phase measurements.  

\section{Results}
\label{sect_res}

\subsection{Continuum data}
\label{sect_res_cont}

We selected the continuum spectral channels shortward of the CO band head 
at 2.295~\mbox{$\mu$m}, avoiding the CO lines as well as the weak lines.  
The average visibilities over the continuum spectral channels were 
computed in each data set.  
We adopted the simple mean of the errors without 
reducing by $\sqrt{N_{\rm cont}}$, where $N_{\rm cont}$ is the number of 
the continuum spectral channels, because the errors are dominated by 
systematic errors, and they do not become smaller by averaging.  

Figure~\ref{vis_continuum} shows the continuum visibilities measured 
in 2009 and 2010.  
The 2009 and 2010 data sample up to the fourth and fifth visibility lobes, 
respectively.  This corresponds to angular resolutions of 9.8 and 7.2~mas 
for the 2009 and 2010 data, respectively, which mark the highest spatial 
resolution achieved for Antares at any wavelength.  
Uniform-disk fits to the 2009 and 2010 data result in $36.97 \pm 0.05$~mas 
(reduced $\chi^2$ = 4.6) and $36.50 \pm 0.05 $ mas (reduced $\chi^2$ = 5.4), 
respectively.  Fitting with a power-law-type limb-darkened disk 
(Hestroffer et al. \cite{hestroffer97}) results in 
a limb-darkened disk diameter of $37.38 \pm 0.06 $~mas and a limb-darkening 
parameter of $(8.7 \pm 1.6) \times 10^{-2}$ for the 
2009 data and $37.31 \pm 0.09$~mas and $(1.5 \pm 0.2) \times 10^{-1}$ for the 
2010 data 
(the errors were estimated by the bootstrapping technique as described in 
Efron \& Tibshirani \cite{efron93}).  

The reduced $\chi^2$ values for the fit to the 2009 and 2010 data are 3.9 and 
3.3, respectively.  
These values are better than those with the uniform-disk fit but are still 
higher than 1.  
This is because of the deviation from the limb-darkened disks in the 
data points at visibility nulls at spatial frequencies of 
$\sim$87~arcsec$^{-1}$ (2009 data) 
and $\sim$115~arcsec$^{-1}$ (2010 data), as well as those at 
95--100~arcsec$^{-1}$ in the 2009 data.  
These last data points were obtained at position angles differing by 
20--25\degr, which suggests the presence of inhomogeneities on a 
spatial scale of $\sim$10~mas in 2009.  
However, as discussed in Sect.~\ref{subsect_imaging}, the reconstructed 
images in the continuum show only a very slight deviation from the 
limb-darkened disk: less than 2\% (see Fig.~\ref{images2009}i).  
The 2010 data include $u\varv$ points taken at position angles differing by 
roughly 90\degr, which are located at spatial frequencies of 
130--138~arcsec$^{-1}$.  
Still, Fig.~\ref{vis_continuum} shows that the measured visibilities closely 
follow the limb-darkened disk.  
In summary,  the overall deviation from the limb-darkened disks is small, 
indicating that the star shows only a weak signature of inhomogeneities in the 
continuum.  
Comparison between the 2009 and 2010 data suggests that a time variation 
with an interval of one year is only marginal.  
The same finding in Betelgeuse is reported in Paper II.  
Therefore, RSGs may show only a 
small deviation and time variation in the surface structure seen in the 
continuum.  

This observed time variation (or absence of it) is much smaller than 
the maximum visibility variation of $\pm$40\% predicted by the current 
3-D convection simulations in the third lobe for 2.2~\mbox{$\mu$m}, which 
approximately samples the continuum (see Fig.~18 of 
Chiavassa et al. \cite{chiavassa09}).  
However, this does not necessarily mean that the observed time variation 
disagrees with the 3-D simulations.  It is possible that we observed 
Antares at two epochs with small variations by chance, given that 
the 3-D simulations of Chiavassa et al. (\cite{chiavassa09}) 
show that large convective cells have lifetimes of years.  
Furthermore, the standard deviation of the temporal variation 
in the visibility predicted by the 3-D simulations 
is smaller than 40\%, down to 10\%, depending on the spatial frequency 
(see Figs.~11, 12, and 14 of Chiavassa et al. \cite{chiavassa09}).  
It is also possible that significant inhomogeneities remained 
undetected due to the limited position angle coverage of our observations. 
Therefore, long-term monitoring AMBER observations with better position 
angle coverage are necessary to rigorously test the current 3-D convection 
simulations. 

The derived limb-darkened disk angular diameters are noticeably smaller than 
the $41.3 \pm 0.1$~mas 
derived from lunar occultation measurements of Richichi \& Lisi 
(\cite{richichi90}).  
However, they used a narrow-band filter at 2.43~\mbox{$\mu$m}\ with a FWHM of 
0.035~\mbox{$\mu$m}.  A number of CO first overtone lines and possibly also \mbox{H$_2$O}\ 
lines from the MOLsphere are present in this wavelength range, which makes 
the star appear larger than in the continuum.  
Furthermore, Richichi \& Lisi (\cite{richichi90}) detected asymmetry 
in the brightness profile reconstructed from the lunar occultation data. 
This also indicates that their angular diameter is affected by the 
inhomogeneous outer atmosphere, where the CO and \mbox{H$_2$O}\ molecules form.

\subsection{Determination of basic stellar parameters}
\label{sect_res_param}

\begin{figure}
\resizebox{\hsize}{!}{\rotatebox{0}{\includegraphics{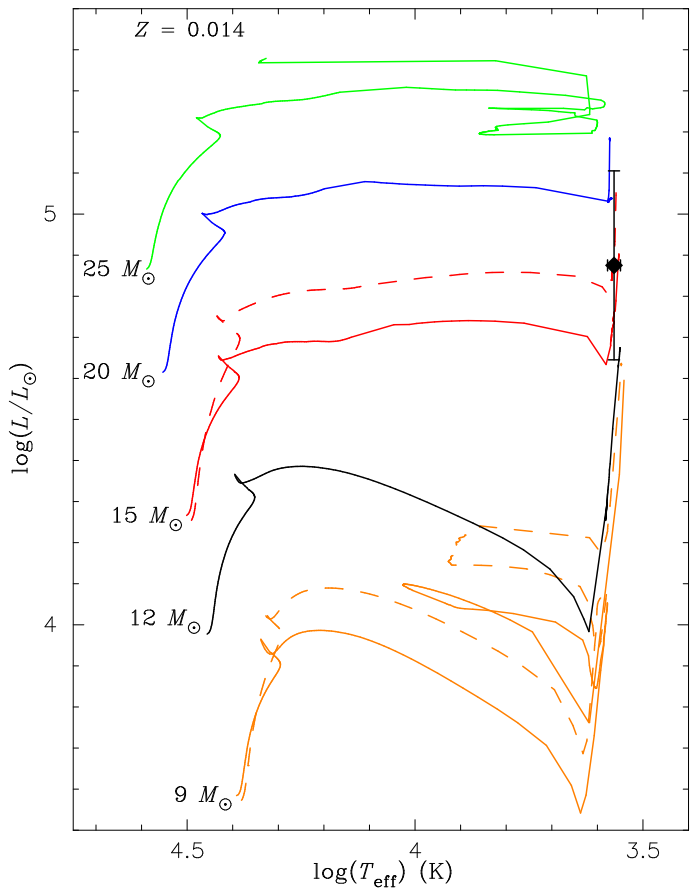}}}
\caption{
H-R diagram with the theoretical evolutionary tracks for 9, 12, 15, 20, and 
25~\mbox{$M_{\sun}$}\ stars for $Z$=0.014 without rotation (solid lines) 
from Ekstr\"om et al. (\cite{ekstroem12}) 
and the observationally derived position of Antares (filled diamond with 
error bars).  For the 15 and 9~\mbox{$M_{\sun}$}\ stars, the evolutionary tracks with 
rotation are plotted by the dashed lines.  
}
\label{hr_diagram}
\end{figure}

By combining the above angular diameters in the continuum with the measured 
bolometric flux, we can determine the effective temperature.  
We collected photometric data from the $U$ band to 10~\mbox{$\mu$m}\ from the 
literature (Lee \cite{lee70}; Ducati \cite{ducati02}; 2MASS, Skrutskie 
et al. \cite{skrutskie06}).  The photometric data were dereddened by 
the interstellar extinction, which was estimated as follows.  The intrinsic 
$B-V$ color of a red supergiant with the spectral type of Antares (M1.5Iab-b) 
is 1.70 (Elias et al. \cite{elias85}).  On the other hand, the observed $B-V$ 
color of Antares is 1.84 (Ducati \cite{ducati02}).  
Therefore, the color excess $E(B-V)$ due to the 
interstellar extinction is 0.14, which translates into $A_{V}$ = 0.43 if 
$A_{V} = 3.1 E(B-V)$ (Savage \& Mathis \cite{savage79}) is assumed.  
Using this interstellar extinction, 
we derived a dereddened bolometric flux 
of $8.36\times10^{-8}$~W~m$^{-2}$.  
The irregular variability of Antares with an amplitude of $\sim$1 magnitude 
in the visible leads to an uncertainty in the bolometric flux, because the 
photometric data collected from the literature were not taken simultaneously. 
However, the time variation in flux between the $R$ and $K$ bands is difficult 
to estimate from the data available in the literature, because the data points 
are not sufficient.  Therefore, we checked the time variation in flux in these 
bands for Betelgeuse and adopted it for Antares.  
We estimated the amplitude of the time variation in the $R$- and $I$-band 
flux of Betelgeuse to be 15\% based on the photometric data from Ducati
(\cite{ducati02}) and Low et al. (\cite{low70}).  The light curves at 1.25, 
2.2, 3.5, and 4.9~\mbox{$\mu$m}\ obtained over 3.6 years with COBE/DIRBIE suggest 
that the amplitude of the flux variation is approximately 10\% in these IR 
bands (Price et al. \cite{price10}). 
With these flux variations adopted for Antares, the uncertainty in the 
bolometric flux is estimated to be $\pm 1.07\times10^{-8}$~W~m$^{-2}$.  

The limb-darkened disk diameters derived from our AMBER continuum data 
and this bolometric flux result in an effective temperature of $3660 \pm 
120$~K for both 2009 and 2010.  
This effective temperature agrees well with the effective temperature scale 
of Levesque et al. (\cite{levesque05}), which gives 3710~K for the spectral 
type of Antares (M1.5).  
By combining the measured bolometric flux and the parallax of 
$5.89 \pm 1.00$~mas (van Leeuwen \cite{vanleeuwen07}), we derive a luminosity 
of $\log L_{\rm \star}/\mbox{$L_{\sun}$} = 4.88 \pm 0.23$. 
The linear radius estimated from the measured 
limb-darkened disk angular diameters and the parallax is 
680~\mbox{$R_{\sun}$}\ (3.2~AU).  
The effective temperature and luminosity we derived agree very well 
with \mbox{$T_{\rm eff}$}\ = $3707\pm77$~K and $\log L_{\rm \star}/\mbox{$L_{\sun}$} = 4.99\pm0.15$ 
derived by Pecaut et al. (\cite{pecaut12}).  

The observationally derived location of Antares on the H-R diagram is 
shown in Fig.~\ref{hr_diagram} with the theoretical evolutionary 
tracks for 9, 12, 15, 20, and 25~\mbox{$M_{\sun}$}\ stars for the solar metallicity 
$Z$=0.014 published in Ekstr\"om et al. (\cite{ekstroem12}).  
The location of Antares agrees with the evolutionary tracks for a 
15~\mbox{$M_{\sun}$}\ star.  
However, given the large uncertainty in the luminosity, 
the error in the initial stellar mass is estimated to be $\pm5$~\mbox{$M_{\sun}$}.  
This value agrees reasonably well with the 17.2~\mbox{$M_{\sun}$}\ derived by 
Pecaut et al. (\cite{pecaut12}), who estimated the mass using the 
evolutionary tracks with rotation of Ekstr\"om et al. (\cite{ekstroem12}). 
The theoretical evolutionary tracks of Ekstr\"om et al. (\cite{ekstroem12}) 
shows that when a 15~\mbox{$M_{\sun}$}\ star reaches the observationally derived 
effective temperature and luminosity of Antares, the current mass is 
13--14.3~\mbox{$M_{\sun}$}\ (without rotation) and 11--14.3~\mbox{$M_{\sun}$}\ (with rotation).  
Combining the mass of 11--14.3~\mbox{$M_{\sun}$}\ and the linear radius of 680~\mbox{$R_{\sun}$}, 
we obtain a surface gravity of $\log \varg = -0.1$...$-0.2$ 
($\varg$ is given in units of cm~s$^{-2}$ throughout the paper).

The age of Antares is important for constraining the Upper Scorpius 
OB association, which is still controversial.  
Comparison of the position of Antares on the H-R diagram with the 
evolutionary tracks of Ekstr\"om et al. (\cite{ekstroem12}) for a 
15~\mbox{$M_{\sun}$}\ star with and without rotation suggests an age of 11--15~Myr. 
This agrees very well with the $12^{+3}_{-1}$~Myr derived by Pecaut 
et al. (\cite{pecaut12}), which is not surprising given the good 
agreement for the effective temperature and luminosity between 
their results and ours.  
Pecaut et al. (\cite{pecaut12}) show that this age of Antares is 
consistent with the age of the Upper Scorpius OB association of 
9--13~Myr derived from stars with other spectral types.

\subsection{Reconstructed images}
\label{subsect_imaging}

\begin{figure*}
\resizebox{\hsize}{!}{\rotatebox{-90}{\includegraphics{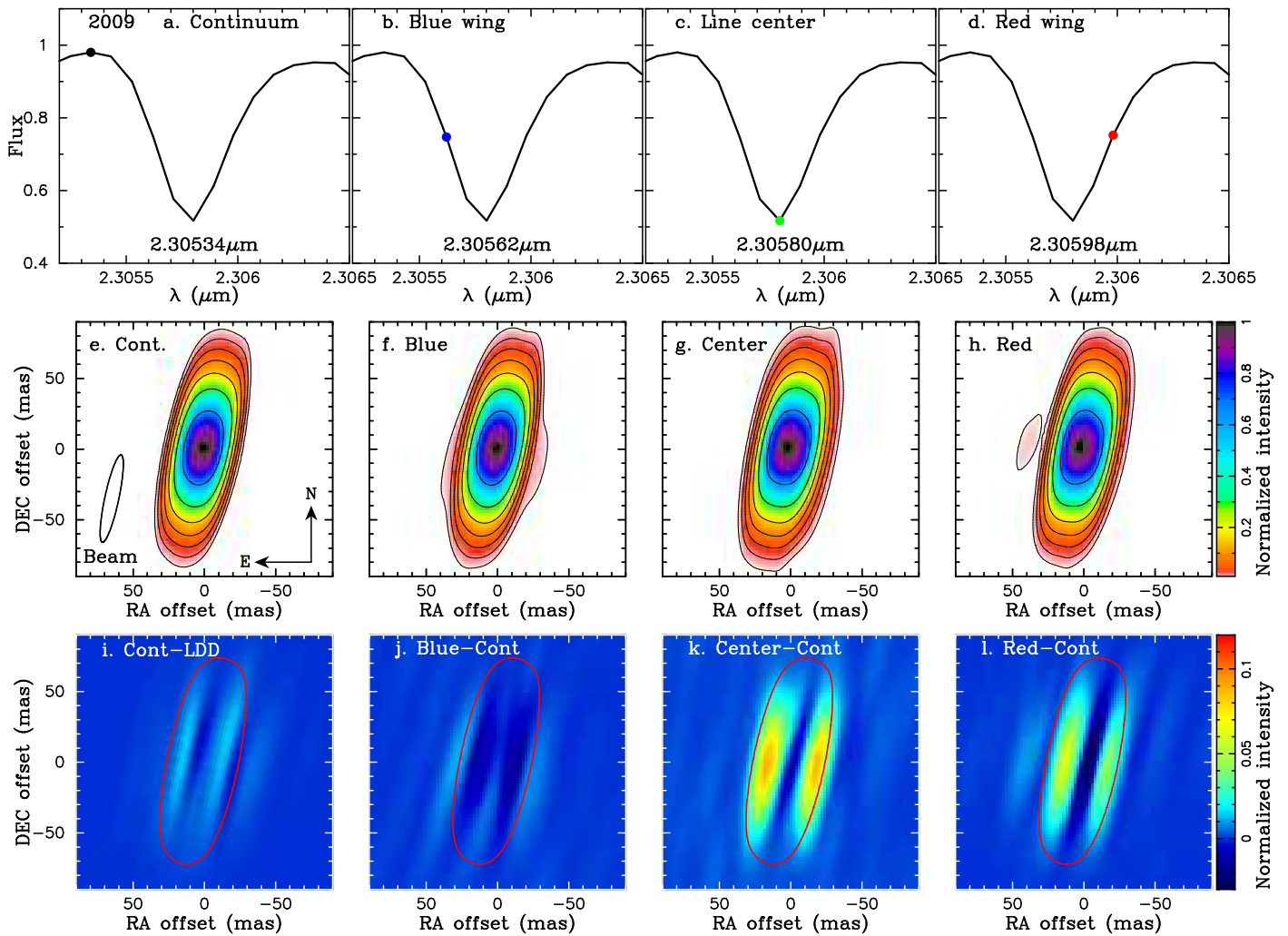}}}
\caption{
Reconstructed images of Antares in 2009 
within one of the CO first overtone lines (2.3058~\mbox{$\mu$m}). 
{\bf a}--{\bf d}: The observed line profile with four representative 
wavelengths marked.  
Panels {\bf e}--{\bf h} show the images reconstructed at these 
wavelengths.  The contours correspond to 1, 2, 4, 8, 16, 32, and 64\% 
of the peak intensity of each image.  
Panel {\bf i} shows the difference between the reconstructed continuum 
image and the best-fit limb-darkened disk for the continuum visibilities, 
whose parameters are given in Sect.~\ref{sect_res_cont}.  
Panels {\bf j}--{\bf l} show the continuum-subtracted images, in which 
the continuum image (panel {\bf e}) is subtracted from the images in 
the panels {\bf f}--{\bf h}.  
The limb of the star as defined in Sect.~\ref{subsect_imaging} is marked by 
the red solid lines.  
}
\label{images2009}
\end{figure*}

Figure~\ref{images2009} shows the reconstructed images from the 2009 data 
at four different wavelengths (panels a--d) 
within the CO line observed at 2.3058~\mbox{$\mu$m} \footnote{A movie of the 
data cube containing all wavelengths from 2.28 and 2.31~\mbox{$\mu$m}\ is 
available at \\http://www.mpifr-bonn.mpg.de/staff/kohnaka}.  
All reconstructed images appear strongly elongated 
due to the elongated beam ($9.8\times63.2$~mas), which results from the 
lack of $u\varv$ points roughly in the North-South direction 
(see Fig.~\ref{uv_coverage}).  
The reconstructed continuum image (Fig.~\ref{images2009}e) shows only a slight 
deviation from the limb-darkened disk that represents the best fit to the 
continuum visibilities.  
This is clearly seen in Fig.~\ref{images2009}i, which shows the residual 
continuum image obtained 
after subtracting the limb-darkened disk with the parameters derived in 
Sect.~\ref{sect_res_cont} convolved with the interferometer's beam 
(the determination of the beam is described in 
Appendix~\ref{appendix_simtests}).  
The residuals are at most 2\% of the peak intensity of the reconstructed 
continuum image.  

The image in the blue wing (Fig.~\ref{images2009}f) appears very similar to 
that in the continuum 
except for the very weak extended component on either side of the star. 
However, the intensity of this extended component is comparable to 
the noise level of the image reconstruction ($\sim$1\%), 
which is estimated from the intensity of the background features 
sufficiently away from the star in the reconstructed images.  
Therefore, 
it is not clear whether or not this extended component in the blue wing image 
is real. 
On the other hand, the images in the line center and red wing 
(Figs.~\ref{images2009}g and \ref{images2009}h) show a much more 
prominent, asymmetrically extended component.  

\begin{figure}
\resizebox{\hsize}{!}{\rotatebox{0}{\includegraphics{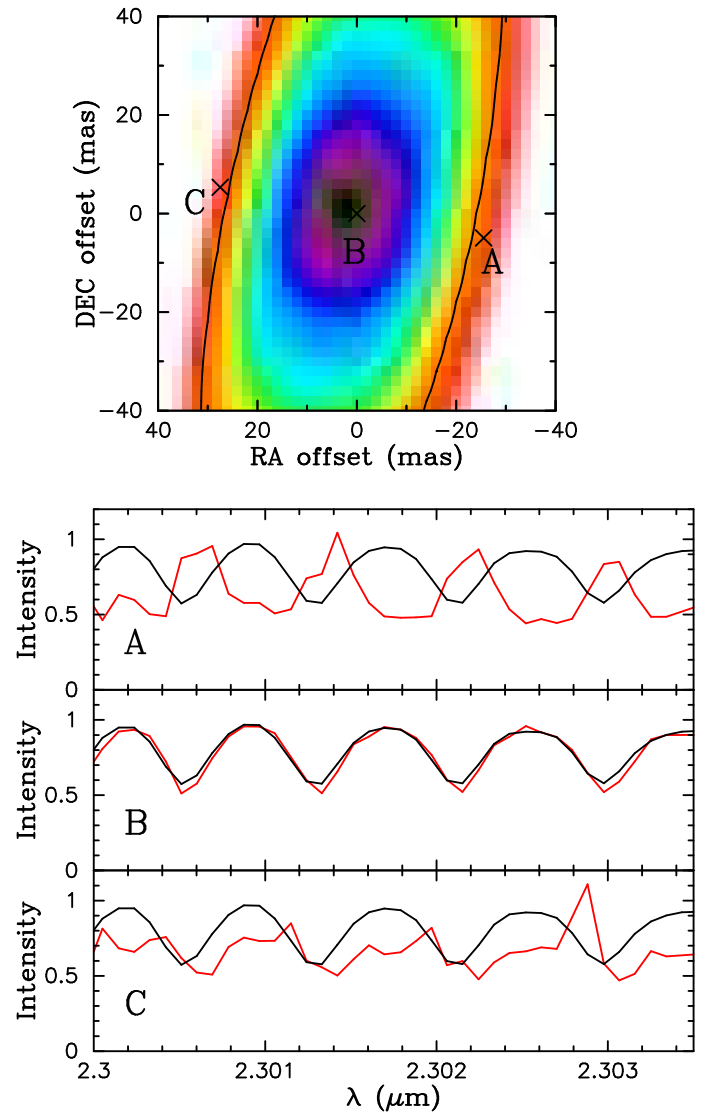}}}
\caption{
Spatially resolved spectra obtained in 2009 at three representative positions 
over the stellar image are plotted by the red solid lines.  
The representative positions are marked in the top panel, which shows the 
line center 
image at 2.30580~\mbox{$\mu$m}\ (same as Fig.~\ref{images2009}g).  
The black lines represent the spatially unresolved spectrum.  
The spatially resolved spectra are scaled for comparison with the spatially 
unresolved spectrum.  The black solid line in the top panel marks the 
limb of the star.  
}
\label{spec2009}
\end{figure}

In order to clearly show the difference among the images, we 
normalized each image with its peak intensity and subtracted the 
normalized continuum image from the images within the line.  
These continuum-subtracted images are shown in the bottom row of 
Fig.~\ref{images2009}.  Also plotted is the limb of the star (red solid line), 
which we define by the contour at 3\% of the peak 
intensity of the continuum image.  
We adopted this contour 
because the CO lines appear in emission outside the limb defined in this 
manner as described below (Fig.~\ref{spec2009}).  
Note that we do not artificially adjust the position of the images 
(either in the lines or in the continuum) at all, 
because the relative astrometry 
among the images at different spectral channels is conserved thanks to 
the self-calibration technique (see Paper II for details). 
The continuum-subtracted images in the line center and red wing clearly 
show an extended component that is more prominent on the western
side of the star, while the continuum-subtracted image in the blue wing shows 
almost no extended component.  
The resolved extended component is present off the limb of the star 
(i.e., outside the stellar disk) 
particularly on the western side of the line center image and also--albeit 
weaker--in the red wing image.  
This represents the second imaging of the MOLsphere in RSGs within 
individual CO lines after Betelgeuse reported in Paper II.  

While we normalized each of the above images with its peak intensity, 
it is also possible to normalize each image so that the intensity integrated 
over the entire image is equal to the observed flux at that spectral channel.  
This allows us to extract the spatially resolved spectrum at each position 
over the stellar image.  
The spatially resolved 2009 spectra at three representative positions along the 
direction of the beam's minor axis are shown in Fig.~\ref{spec2009}, 
which reveals the CO emission lines off the limb on the western side 
(position A)\footnote{A movie of the 
spatially resolved spectra across the stellar image is 
available at \\http://www.mpifr-bonn.mpg.de/staff/kohnaka}.  
These emission lines are redshifted with respect to the absorption lines 
in the spatially unresolved observed spectrum, reflecting the appearance 
of the extended component only in the line center and red wing. 
On the eastern side, the line profiles are characterized by weak 
blueshifted emission and redshifted absorption.  
This is explained as follows.  
Because of the finite beam size, we see not only the emission off the limb 
but also the absorption originating from inside the limb.  
The weak blueshifted emission, which is seen as the weak extended feature 
in the reconstructed blue wing image (Fig.~\ref{images2009}f), partially 
fills in the absorption.  This results in what appears as redshifted 
absorption with weak blueshifted emission.  
Ohnaka (\cite{ohnaka13}) presents similar spatially resolved CO first 
overtone line spectra for Betelgeuse and reveals the CO line emission off 
the limb, but for the 1-D image. 
Our 2-D image reconstruction for Antares demonstrates that it is now 
feasible to obtain the spatially resolved spectrum at each position 
over the surface of a star as well as off the limb of the star, 
as is routinely done in solar physics.  

\begin{figure*}
\resizebox{\hsize}{!}{\rotatebox{-90}{\includegraphics{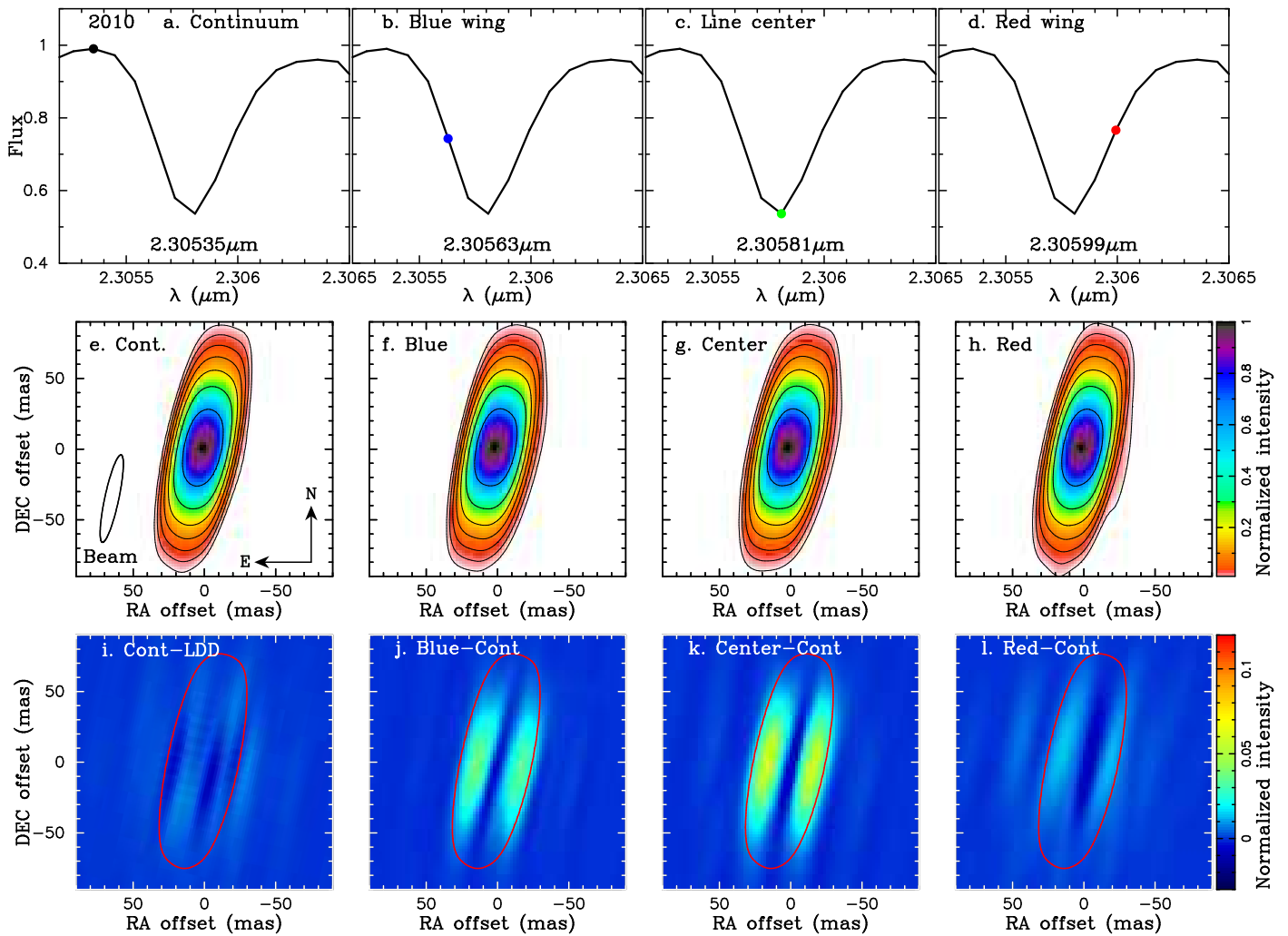}}}
\caption{
Reconstructed images of Antares in 2010 shown in the same manner as 
Fig.~\ref{images2009}.  
}
\label{images2010}
\end{figure*}

Figure~\ref{images2010} shows the reconstructed images from the 2010 data 
for the same CO line as shown in Fig.~\ref{images2009}.  
Because the position angle coverage of the 2010 observations is better than 
the 2009 observations, the beam of the 2010 images is less elongated than 
that of the 2009 images.  However, for a direct comparison of the images 
taken at two epochs, the 2010 images are 
convolved with the same beam as used for the images in 2009. 
The reconstructed image in the continuum shows little deviation from 
the best-fit limb-darkened disk derived from the continuum visibilities, 
as in the 2009 data.  
The residual after subtracting the limb-darkened disk with the parameters 
derived in Sect.~\ref{sect_res_cont} is at most 1\% of the peak intensity 
of the reconstructed continuum image. 
The continuum-subtracted images shown in the bottom row of 
Fig.~\ref{images2010} reveal that the images in the blue wing and 
line center show a prominent extended component.  
On the other hand, the image in the red wing shows little trace of this 
(the intensity of the extended component in the red wing image is $\sim$2\%, 
which is only slightly higher than the image reconstruction noise of 1\%).  
This means that the appearance of the star in the blue and red wing has 
swapped within one year (compare the images in the bottom row of 
Figs.~\ref{images2009} and \ref{images2010}). 

\begin{figure}
\resizebox{\hsize}{!}{\rotatebox{0}{\includegraphics{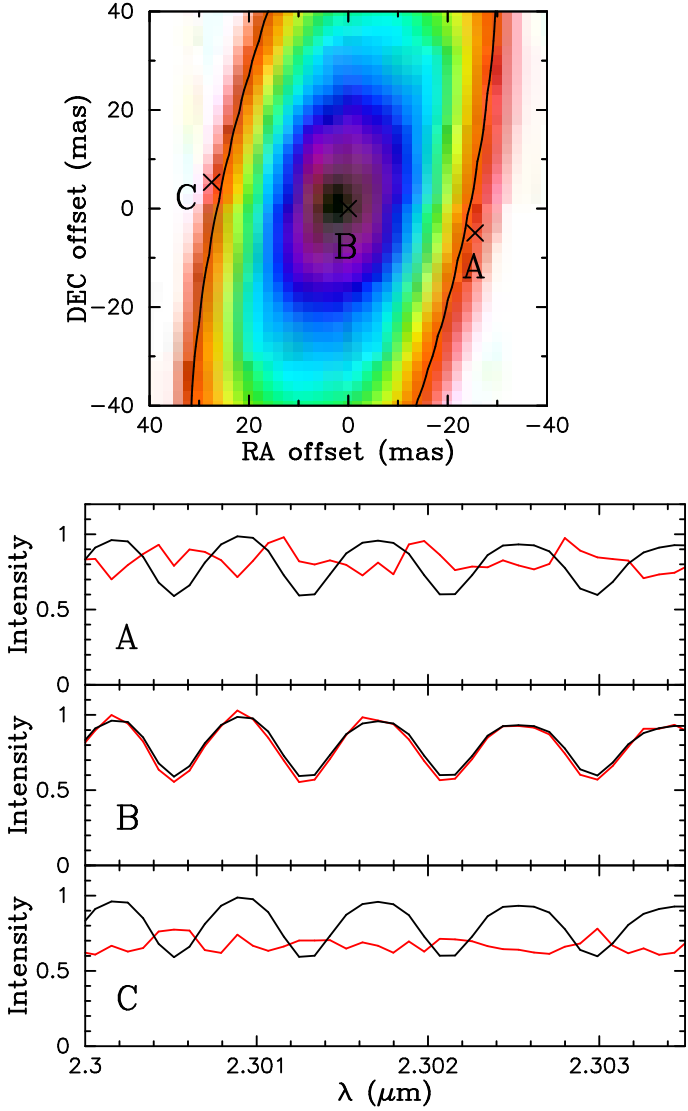}}}
\caption{
Spatially resolved spectra obtained in 2010 shown in the same manner as 
Fig.~\ref{spec2009}.  
}
\label{spec2010}
\end{figure}

\begin{figure}
\resizebox{\hsize}{!}{\rotatebox{-90}{\includegraphics{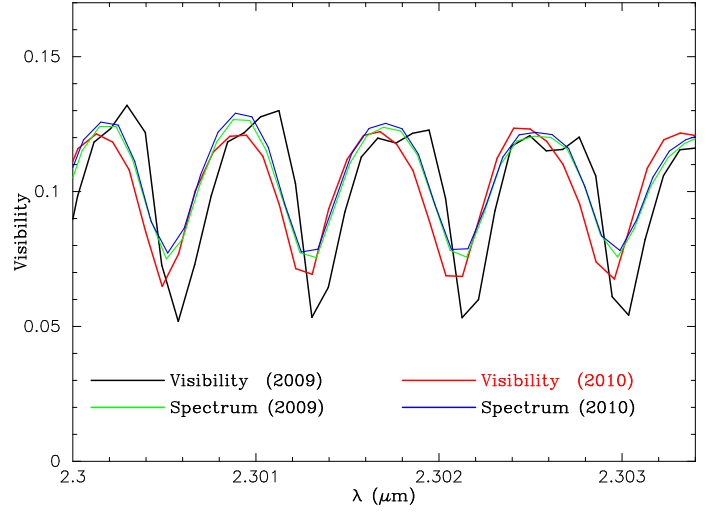}}}
\caption{
Comparison of the visibilities observed in 2009 (data set \#26, black 
solid line) and 2010 (data set \#107, red solid line).  
Both visibilities were taken at almost the same projected 
baseline length (13.9~m) and position angle (84\degr --85\degr ). 
The spatially unresolved spectra observed in 2009 and 2010 are also plotted 
by the green and blue solid lines.  
}
\label{vis2009_2010}
\end{figure}

Figure~\ref{spec2010} shows the spatially resolved spectra at three selected 
positions over the stellar image.  
Obviously, the blueshifted emission appears off the limb (position A), 
although the strength of the emission is weaker than the redshifted 
emission seen in 2009.  This is simply because the extended component, 
MOLsphere, is weaker in 2010 than in 2009.  
This can also be seen in the visibilities 
observed in the first lobe, as shown in Fig.~\ref{vis2009_2010}.  
While the observed spatially unresolved spectra show little 
time variation, the visibilities in the CO lines observed in 2010 are 
noticeably higher than observed in 2009.  This means that the star appears 
less extended in the CO lines in 2010 than in 2009.  
The figure also shows the time variation in the visibilities within the 
CO lines, which corresponds to the different appearance of the star 
within the CO lines as shown in Figs.~\ref{images2009} and \ref{images2010}. 

The time variation in the reconstructed images within the CO lines 
suggests a significant change in the dynamics of the atmosphere. 
Such a time variation in the atmospheric dynamics has already been 
detected in Betelgeuse (Paper II).  Therefore, our AMBER observations of 
Antares implies that the inhomogeneous, highly temporally variable 
nature of the outer atmosphere might be common among RSGs.

\section{Modeling of the velocity field}
\label{sect_modeling}

The different appearance of the star within the individual CO lines was also 
detected in Betelgeuse, and it was interpreted as inhomogeneous gas 
motions in the photosphere and MOLsphere (Papers I and II).  
To characterize the velocity field using the above reconstructed images of 
Antares, we applied the same two-layer model as used for Betelgeuse 
(see Paper I for details of the model).   
The star is represented with the blackbody of 3700~K, which is the effective 
temperature of Antares derived from our continuum visibilities 
(Sect.~\ref{sect_res_param}).  
We represent the CO gas in the photosphere and MOLsphere with 
an inner and outer layer, respectively.  
The reason we adopted this two-layer model is that we cannot reproduce the 
observed CO line spectrum and the CO line images simultaneously with a single 
layer model.  
For example, 
single layer models with CO column densities of 
$10^{21}$--$10^{22}$~\mbox{cm$^{-2}$}\ and a radius of $\sim \!\! 1.5$~\mbox{$R_{\star}$}\ can 
roughly explain the observed extended 
emission in the CO lines.  However, these models predict the CO absorption 
spectrum to be too weak compared to the observations or to appear even in 
emission, because of the strong extended emission from the layer.  
This suggests that the inner layer representing the geometrically compact 
photosphere is necessary to explain the images and spectrum simultaneously.  

Each layer is characterized by its radius, CO column density, and 
temperature.  
As in Papers I and II, the layers are assumed to be geometrically thin, 
and the thickness of the layers are fixed to 0.02~\mbox{$R_{\star}$}.  
Because this choice is arbitrary, 
we computed models with larger geometrical thicknesses 
(up to 0.05 and 0.2~\mbox{$R_{\star}$}\ for the inner and outer layer, respectively) 
and confirmed that the thickness of the layers does not affect the results of 
the modeling.  
Changing the inner layer radius affects the intensity distribution only 
slightly.  The spatial resolution of the current AMBER data is sufficient for 
constraining the size the MOLsphere (i.e., outer radius) but insufficient for 
constraining the inner radius.  
We approximate an inhomogeneous 
velocity field with a patch (or clump) of CO gas that is moving 
at a different velocity from the gas in the remaining region.  

For the inner CO layer, we adopted the same parameters as for Betelgeuse, 
because the effective temperature and surface gravity of Antares are 
similar to those of Betelgeuse (\mbox{$T_{\rm eff}$}\ = 3690~K from Paper II and 
$\log \varg = -0.3$ from the mass and radius estimated by 
Harper et al. \cite{harper08}).  
We set the radius, temperature, and CO column density of the inner 
layer to be 1.05~\mbox{$R_{\star}$}, 2250~K, and $5 \times 10^{22}$~\mbox{cm$^{-2}$}, 
respectively, as in Papers I and II.  
The radius, temperature, and CO column density of the outer layer, 
as well as the velocity field, were treated as free parameters, 
and we assumed the same velocity field for the inner and outer layers.
Because the elongated beam of our AMBER observations makes it difficult to 
know the actual number and shape of stellar spots and the spatial 
resolution in the North-South direction is low, 
we assumed only a single patch of CO gas to keep the number of free 
parameters as small as possible.  
The model images were spectrally convolved with the spectral resolution 
of the observed data ($\lambda / \Delta \lambda = 8000$) and then 
convolved with the same beam as for the reconstructed images.  

\begin{figure*}
\resizebox{\hsize}{!}{\rotatebox{-90}{\includegraphics{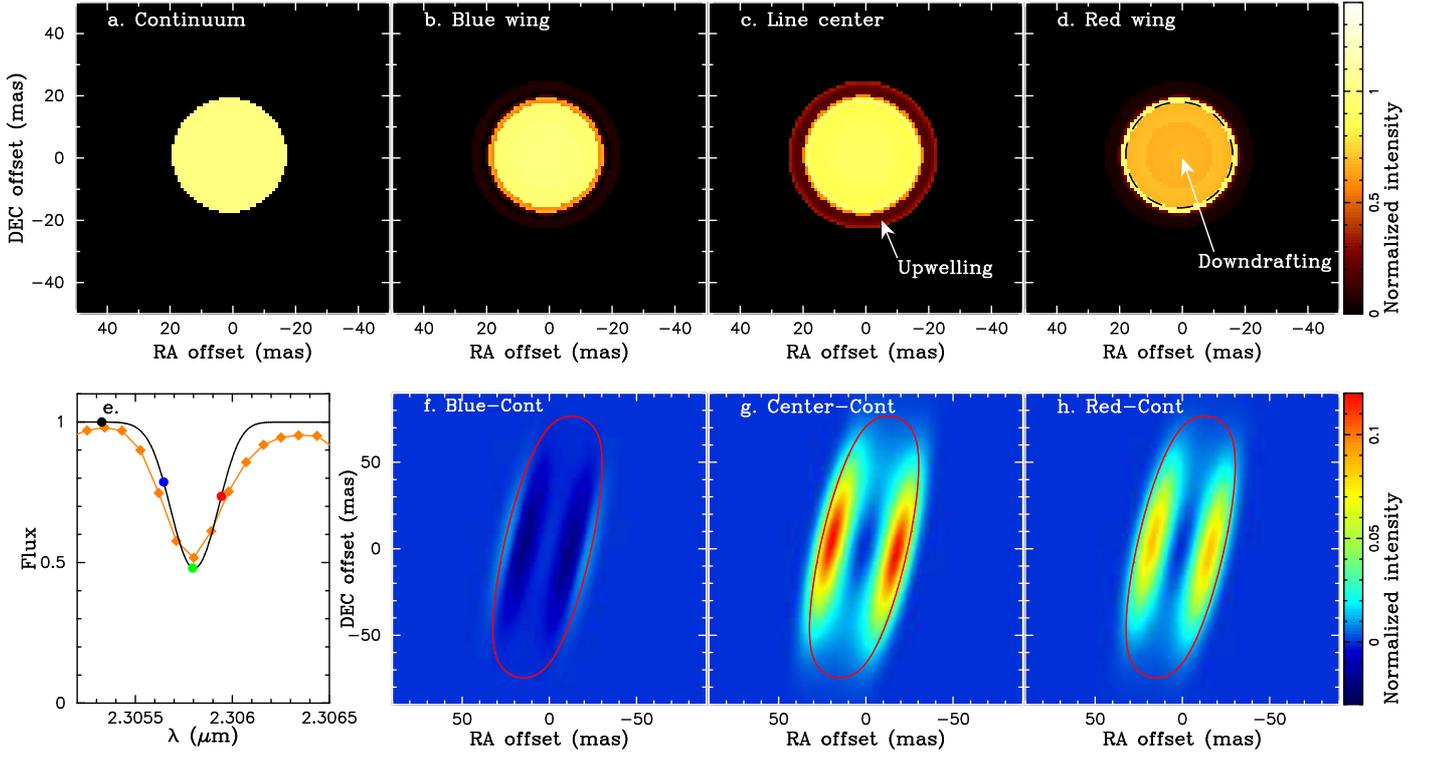}}}
\caption{
Best-fit stellar patch model with an inhomogeneous velocity field for Antares 
observed in 2009.  Panels {\bf a}--{\bf d} show model images in the 
continuum, blue wing, line center, and red wing within the same CO line 
profile as shown in Fig.~\ref{images2009}.  
The wavelengths of these images are marked with black, blue, green, and 
red dots in 
Panel {\bf e}, where the observed and model spectra are plotted by the 
filled diamonds and the black solid line, respectively.  
Panels {\bf f}-{\bf h} show the best-fit continuum-subtracted model images 
convolved with the same beam as the observed images.   Therefore, these 
model images can be compared to the images shown in 
Figs.~\ref{images2009}j--\ref{images2009}l.   
}
\label{model2009}
\end{figure*}

Figures~\ref{model2009}f--h show the best-fit continuum-subtracted model 
images derived for the 2009 observations 
within the same CO line as shown in Fig.~\ref{images2009}. 
A comparison between the observed line profile and that predicted by the 
model is also shown.  
This model is characterized by a single large patch (or clump) of CO gas 
infalling with 2.5~\mbox{km s$^{-1}$}, which is marked with the dashed circle in the 
red wing image in Fig.~\ref{model2009}d (see also the schematic view in 
Fig.~\ref{schematic}). 
This CO gas patch or clump produces slightly redshifted 
absorption, which explains why the intensity within the patch in the 
red wing image (Fig.~\ref{model2009}d) is lower than in the remaining region.  
The CO gas outside the patch is moving outward much faster at 25~\mbox{km s$^{-1}$}, 
as marked in the line center image in Fig.~\ref{model2009}c 
(see also the schematic view in Fig.~\ref{schematic}).  
The outer CO layer extends to 1.3~\mbox{$R_{\star}$}\ with a temperature of 2000~K and 
a CO column density of $1\times 10^{20}$~\mbox{cm$^{-2}$}.  
The extended component observed in the line center and red wing, as well as 
the absence of it in the blue wing, is reasonably reproduced.  
From the observed images, the velocity of the CO gas moving inward and 
outward is constrained to be 0--5~\mbox{km s$^{-1}$}\ and 20--30~\mbox{km s$^{-1}$}, respectively. 
However, the model predicts the extended component to be too strong, 
and the observed asymmetry of the extended component (i.e., it extends 
only to the western side) is not reproduced.  The model flux is also 
too narrow compared to the observed line profile.

A summary of our modeling is given in Table~\ref{table_param}.  
We also computed the reduced $\chi^2$ value for each model in the continuum, 
blue wing, line center, and red wing from the fit to the observed visibilities 
and closure phases.  
The aforementioned differences between the model and the observed images 
are reflected in the reduced $\chi^2$ values that are much higher than 1. 
This is presumably 
due to the simplifications adopted in our model.  In particular, we assumed 
the (radial) column density of CO and temperature in each layer 
to be constant regardless of the direction with respect to the stellar center 
and attempted to explain the observed results only by introducing a 
simplified velocity field.  We also assumed a single patch to keep the 
number of free parameters as small as possible.  
Therefore, inhomogeneities in the CO density and temperature, as well as 
multiple patches, may reconcile the disagreement between the 
observed data and the model.  However, our goal is to obtain an approximate 
picture of the velocity field, and therefore, we refrain from introducing 
more free parameters, which cannot be well constrained from the present 
data.

\begin{figure*}
\resizebox{\hsize}{!}{\rotatebox{-90}{\includegraphics{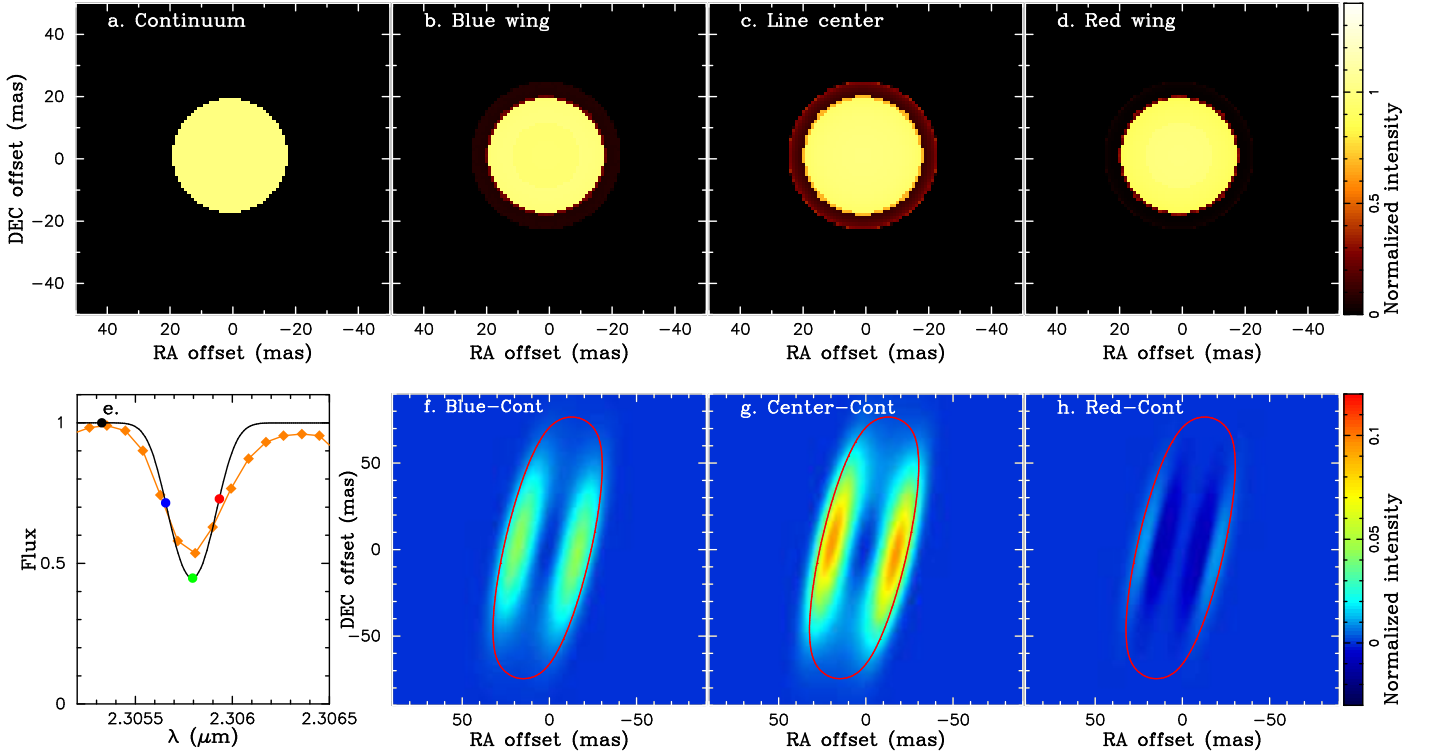}}}
\caption{
Best-fit stellar patch model with an inhomogeneous velocity field for Antares 
observed in 2010, shown in the same manner as Fig.~\ref{model2009}.  
The continuum-subtracted images from the best-fit model, which are shown in 
Panels {\bf f}-{\bf h}, are compared to the images shown in 
Figs.~\ref{images2010}j--\ref{images2010}l.   
}
\label{model2010}
\end{figure*}

\begin{figure}
\resizebox{\hsize}{!}{\rotatebox{0}{\includegraphics{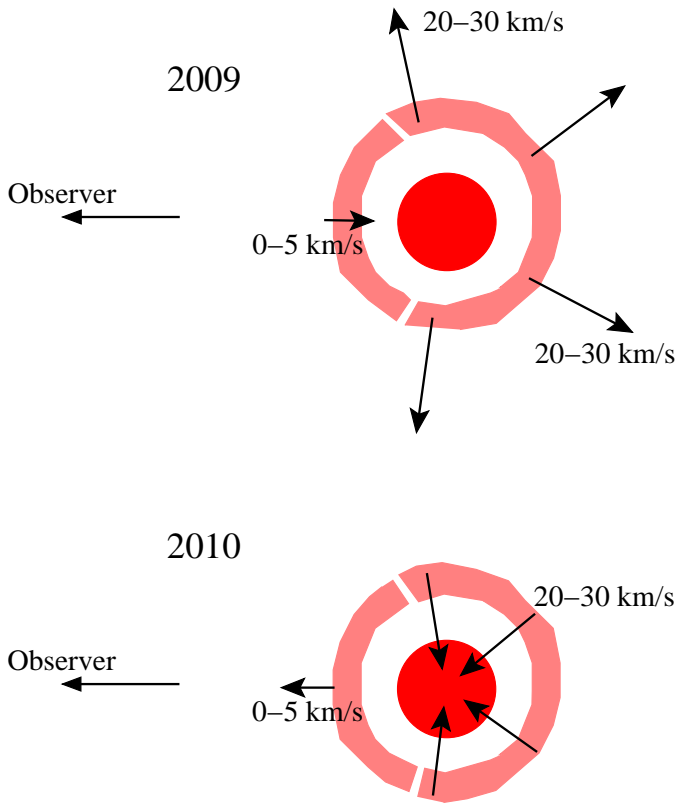}}}
\caption{
Schematic view of the velocity fields in Antares derived from our AMBER 
observations in 2009 and 2010.  Only the outer CO layer is drawn for the 
sake of clarity of the figure, but the velocity field of the inner CO layer 
was assumed to be the same as in the outer layer.  
The images are not to scale. 
}
\label{schematic}
\end{figure}

The best-fit continuum-subtracted model images for the 2010 data 
are shown in Fig.~\ref{model2010}.  In this model, a large patch of CO gas 
is moving outward only with 2.5~\mbox{km s$^{-1}$}, while the CO gas in the remaining 
region is infalling much faster with 25~\mbox{km s$^{-1}$}\ 
(see also the schematic view in Fig.~\ref{schematic}). 
The patch is so large that it dominates the surface of the star facing 
toward the observer.  
The radius and temperature of the outer CO layer are the same as those 
for the 2009 data, but the CO column density ($5\times 10^{19}$~\mbox{cm$^{-2}$}) 
is smaller than that derived from the 2009 data.  
The extended component observed in the line center and blue wing is fairly 
reproduced, and the model image in the red wing shows little trace 
of the extended component as observed.  
The model predicts the extended component to be too strong compared to the 
observed image, and the line profile predicted by the model is also too 
shallow.  However, as mentioned above, this disagreement is likely due to 
the simplifications adopted in our model.  
The smaller CO column density of the outer CO layer found for the 2010 
data reflects the weaker extended component in 2010 compared to 
2009. 
However, we note that this weaker extended component can also be explained 
by a model with a smaller radius of 1.2~\mbox{$R_{\star}$}\ and a CO column density 
of $1\times 10^{20}$~\mbox{cm$^{-2}$}\ for the outer CO layer. 
We also found that the models with the upwelling CO gas with 0--5~\mbox{km s$^{-1}$}\ 
and downdrafting gas with 20--30~\mbox{km s$^{-1}$}\ can explain the observed data.  
The uncertainties in the radius, temperature, and CO column density of 
the outer CO layer are estimated to be $\pm$0.1~\mbox{$R_{\star}$}, $\pm$200~K, and 
a factor of 2, respectively.  

Comparison between the velocity field obtained from the 2009 and 2010 data 
reveals that the overall gas motions in the outer atmosphere changed the 
direction within one year: the strong upwelling motions were dominant in 
2009, while the velocity field in 2010 was dominated by strong downdrafts.

\begin{table*}
\begin{center}
\caption {Parameters of the MOLsphere model with an inhomogeneous velocity
  field for Antares.  The definition of the patch size and position is 
  depicted in Fig.~4 of Paper~I.}
\vspace*{-2mm}

\begin{tabular}{l l l}\hline
Parameter &  Value (2009) & Value (2010) \\
\hline
Inner layer &   & \\
\hline
CO column density (\mbox{cm$^{-2}$}) & $5\times10^{22}$ (fixed) & 
$5\times10^{22}$ (fixed) \\
Temperature (K) & 2250  (fixed) & 2250  (fixed) \\
Radius (\mbox{$R_{\star}$}) & 1.05  (fixed) & 1.05  (fixed)\\
\hline
Outer layer & & \\
\hline
CO column density (\mbox{cm$^{-2}$}) & 
$1\times10^{20}$ ($\pm 0.3$~dex) & $5\times10^{19}$ ($\pm 0.3$~dex) \\
Temperature (K) &
$2000\pm200$ & $2000\pm200$ \\
Radius (\mbox{$R_{\star}$}) & $1.3\pm0.1$ & $1.3\pm0.1$ \\
\hline
Velocity field & &  \\
\hline
Velocity within the patch (\mbox{km s$^{-1}$}) & $2.5\pm2.5$ (inward) & $2.5\pm2.5$ (outward) \\
Velocity outside the patch (\mbox{km s$^{-1}$}) & $25\pm5$ (outward) & $25\pm5$ (inward) \\
Stellar patch size ($\Theta$) (\degr) & $60\pm10$ & $80\pm10$ \\
Stellar patch position ($\theta$, $\phi$) (\degr) & (0--30, 0--360) & (0--30, 0--360) \\
\hline
Reduced $\chi^2$ &  & \\
\hline
Continuum & 10.6 & 18.1 \\
Blue wing & 32.6 &  31.4 \\
Line center & 21.0 & 18.9 \\
Red wing  &  29.8  & 20.5  \\
\hline
\label{table_param}
\vspace*{-7mm}
\end{tabular}
\end{center}
\end{table*}

\section{Discussion}
\label{sect_discuss}

The characteristics of the velocity field found in Antares are similar to 
those reported for Betelgeuse in Papers I and II.  In both stars, 
we found CO gas motions with velocity amplitudes of up to 
20--30~\mbox{km s$^{-1}$}\ on a spatial scale larger than the stellar angular radius.  
Significant time variations in the velocity field have been detected 
over an interval of one year. 
The radius, temperature, and density of the outer CO layer derived for 
Antares are also similar to those in Betelgeuse.  
Therefore, these inhomogeneous velocity fields in the MOLsphere 
extending to 1.2--1.5~\mbox{$R_{\star}$}\ might be common among RSGs.  

A schematic view of the derived velocity fields in 2009 and 2010 is 
given in Fig.~\ref{schematic}.  
It is not yet known whether all outwardly moving gas clumps fall back 
after some time or they are further accelerated outward.  
If we assume that the inhomogeneous CO gas motions in Antares are ballistic, 
the CO gas moving outward with 20--30~\mbox{km s$^{-1}$}\ seen at 1.3--1.4~\mbox{$R_{\star}$}\ in 
April 2009 reaches a maximum height of 1.45--1.60~\mbox{$R_{\star}$}\ in 2--3~months 
and falls back to the original height in another 2--3~months.  
Likewise, if we assume that the CO gas infalling fast at 20--30~\mbox{km s$^{-1}$}\ seen 
in 2010 is falling back from some height with an initial velocity of zero, 
it is estimated to have traveled from 1.45--1.60~\mbox{$R_{\star}$}\ over 2--3~months. 
On the other hand, the CO gas moving inward at only 0--5~\mbox{km s$^{-1}$}\ 
(2009) or outward (2010) may represent a clump that has traveled from a 
deeper layer and is near the maximum height, at which the velocity is 
zero.  In this case, the clump is estimated to have traveled from 
$\sim$1.2~\mbox{$R_{\star}$}\ with an initial velocity of 20--30~\mbox{km s$^{-1}$}\ over 
approximately two months. 
This means that multi-epoch AMBER observations over 2--6~months would 
be useful for probing whether the outwardly moving gas clump falls back or 
it is accelerated outward further.

The physical mechanism responsible for the formation of the MOLsphere 
and the inhomogeneous gas motions in this region is still unclear.  
The upwelling and downdrafting motions are reminiscent of convection, 
which may levitate the gas to form the MOLsphere.  
For example, Lim et al. (\cite{lim98}) suggest that convection could 
lift the photospheric material into the extended atmosphere.  
If convection alone is responsible for the formation of the MOLsphere, 
3-D convection simulations should predict the density consistent 
with the observationally derived values at the location of the 
MOLsphere (1.2--1.4~\mbox{$R_{\star}$}).  

Given that the current 3-D convection simulations cover up to 8--9~\mbox{$R_{\star}$}, 
it is possible to carry out this comparison.  
One of the 3-D models of Chiavassa et al. (\cite{chiavassa11}), which 
is closest to Antares or Betelgeuse, shows that the density at 1.2~\mbox{$R_{\star}$}\ 
ranges between $10^{-20}$ and $10^{-25}$~g~cm$^{-3}$ (see their Fig.~4). 
On the other hand, we can estimate the gas density in the MOLsphere as 
follows.  
While we adopted the geometrical thickness of 0.02~\mbox{$R_{\star}$}\ for the 
outer CO layer representing the MOLsphere, the inner radius of the MOLsphere 
is not observationally constrained.  Therefore, we estimated a lower limit 
on the CO density by dividing the CO column density with the geometrical 
thickness defined by the outer radius of the MOLsphere (i.e., geometrical 
thickness = 0.2--0.4~\mbox{$R_{\star}$}). 
If we assume chemical equilibrium in the MOLsphere, we can estimate 
the gas density from the CO density and the temperature of the MOLsphere 
(2000~K).  
The derived gas density is $\sim \!\! 10^{-14}$~\mbox{cm$^{-3}$}, which is higher 
than the values predicted by the 3-D convection model by six to eleven orders 
of magnitude.  
The analysis of line profiles in Antares by Gray \& Pugh (\cite{gray12}) 
also implies that convective cells penetrate only the lower portion of 
the photosphere, although it is not straightforward to convert the 
radial velocities derived from spatially unresolved spectra to the actual 
velocity field.  
Therefore, at the moment, convection alone is unlikely to explain the 
formation of the MOLsphere and the inhomogeneous gas motions. 
However, it remains to be tested whether the combination of convection 
and some other mechanism (e.g., radiation pressure on molecules and atoms 
or scattering on dust grains) can explain the formation of the MOLsphere 
and the mass loss in RSGs.

Another possible mechanism is the propagation of Alfv\'en waves.  
Magnetic fields have been detected in Betelgeuse (Auri\`ere et al. 
\cite{auriere10}).  
While Grunhut et al. (\cite{grunhut10}) detected no magnetic fields in 
Antares, this can be due to the insufficient S/N as they themselves note. 
Therefore, future observations with a better S/N may reveal magnetic 
fields in more RSGs including Antares, and the magnetohydrodynamical 
(MHD) wave propagation remains a 
possible process to explain the formation of the MOLsphere and the 
inhomogeneous gas motions.  
However, the currently available MHD model by Airapetian et al. 
(\cite{airapetian00}) deals with the stellar winds from the chromosphere and 
does not include the MOLsphere.  
The recent magneto-rotator model by Thirumalai \& Heyl (\cite{thirumalai12}) 
represents an interesting alternative for the RSG mass loss, although their 
model is still stationary and does not include the chromosphere.  
The inclusion of a proper treatment of the time-dependent, multicomponent 
nature (chromosphere and MOLsphere) would be crucial for testing whether or 
not the MHD process with or without rotation is responsible for the formation 
of the MOLsphere and the inhomogeneous gas motions. 

While the presence of the MOLsphere is confirmed by the IR interferometric 
imaging of Antares and Betelgeuse in the individual CO lines, 
there are still problems with the MOLsphere, as mentioned in Ohnaka et al. 
(\cite{ohnaka12}).  The additional TiO absorption originating in the 
MOLsphere makes the TiO bands too strong compared to the observed spectra 
of RSGs (Hron et al. \cite{hron10}).  As Ohnaka et al. (\cite{ohnaka12}) 
suggest, the scattering due to TiO molecules may reconcile the above 
problem, because the line formation by scattering can be important 
for electronic transitions such as the TiO bands in the visible. 
Another problem is that the current MOLsphere models cannot explain the 
high resolution 12~\mbox{$\mu$m}\ spectra of \mbox{H$_2$O}\ lines (Ryde et al. 
\cite{ryde06}).  Recently, Lambert et al. (\cite{lambert13}) suggest 
the importance of NLTE effects in the formation of these \mbox{H$_2$O}\ lines. 
Therefore, modeling with the scattering of 
molecules and NLTE effects taken into account is necessary for 
a comprehensive understanding of the MOLsphere.

\section{Concluding remarks}
\label{sect_concl}

We have presented high spectral resolution 
aperture-synthesis imaging of Antares at two 
epochs with VLTI/AMBER.  
With the stellar surface spatially resolved with AMBER's high spatial 
resolution, AMBER's high spectral resolution 
has enabled us to derive the kinematical 
information at different positions over the surface of the star. 
The reconstructed images obtained at the first epoch (April 2009) reveal an 
asymmetrically extended component (MOLsphere) in the red wing and line center 
of the individual lines, while the image in the blue wing shows little trace
of the extended component.  
At the second epoch (April 2010), the extended component appears in 
the blue wing and line center, and the red wing image does not exhibit the 
extended component.  We have also 
found that the extended component became weaker in 2010.  
These results can be interpreted as a significant time variation in the 
atmospheric dynamics.  

Our modeling suggests that the velocity field in the MOLsphere in 2009 is 
characterized 
by a large patch or clump of CO gas infalling at only 0--5~\mbox{km s$^{-1}$}\ 
with the CO gas in the remaining region moving outwardly much faster 
at 20--30~\mbox{km s$^{-1}$}.   The data observed in 2010 are explained by a 
model in which a large patch of CO gas moving outwardly at 0--5~\mbox{km s$^{-1}$}\ and 
the CO gas in the remaining region is infalling faster at 20--30~\mbox{km
  s$^{-1}$}. 
Our modeling also shows that the MOLsphere is extended to 1.2--1.4~\mbox{$R_{\star}$}\ 
with column densities of $5\times10^{19}$--$1\times10^{20}$~\mbox{cm$^{-2}$}\ and 
a temperature of $\sim$2000~K.  
These properties of the MOLsphere of Antares, as well as the velocity field, are 
very similar to those derived for Betelgeuse.  Therefore, the inhomogeneous 
gas motions in the MOLsphere extending to 1.2--1.5~\mbox{$R_{\star}$}\ might be 
common among RSGs.  

The origin of the detected gas motions is not clear yet.  
To understand the physical mechanism responsible for the dynamics and 
formation of the MOLsphere, it is necessary to spatially resolve 
the dynamics at different heights from the deep photosphere to the outer 
atmosphere.  
Similar AMBER observations for different atomic and molecular lines would 
make this possible.  
For example, weak atomic lines 
such as Ti, Na, and Ca would be useful for probing the dynamics in the 
deep photospheric layers.  
Moreover, it is essential to obtain images with better $u\varv$ coverage. 
This is now feasible with more telescope configurations appropriate for 
the aperture-synthesis imaging available at the VLTI.

\begin{acknowledgement}
We thank the ESO VLTI team for supporting our AMBER observations.  
We are also grateful to Eric Thi\'ebaut, who makes his image 
reconstruction software MiRA publicly available.  
NSO/Kitt Peak FTS data on the Earth's telluric features 
were produced by NSF/NOAO.
\end{acknowledgement}

\appendix
\section{Summary of AMBER observations}

Our AMBER observations of Antares are summarized in Tables~\ref{obslog2009} and 
\ref{obslog2010}.

\renewcommand{\tabcolsep}{0.12cm}
\begin{table*}
\begin{center}
\caption {Log of AMBER observations of Antares and the calibrator 
$\alpha$~Cen A in April 2009 with the E0-G0-H0 (16-32-48m) baseline 
configuration.  The seeing ($s$) and the coherence time ($\tau_0$) are in 
the visible.  
}

\begin{tabular}{r l c c l r r | r l c c l r r}\hline
\# & $t_{\rm obs}$ & $B_{\rm p}$ & PA  & $s$ & $\tau_0$ & NDIT &
\# & $t_{\rm obs}$ & $B_{\rm p}$ & PA  & $s$  & $\tau_0$ & NDIT  \\
   &  (UTC)       &      (m)   &(\degr)&  (\arcsec)      & (ms)     &      &
   &  (UTC)       &      (m)   &(\degr)&   (\arcsec)     & (ms)     &        \\
\hline
\multicolumn{14}{c}{Antares} \\
\hline
\multicolumn{7}{c|}{2009 Apr 24} & \multicolumn{7}{c}{2009 Apr 25} \\
\hline
   1 & 05:14:38 & 30.06/15.04/45.09 &  58/ 58/ 58 &  0.65 &  4.2 &   500	&   36 & 05:01:04 & 29.67/14.84/44.51 &  56/ 56/ 56 &  0.52 &  5.0 &   200 \\ 
   2 & 05:16:57 & 30.14/15.08/45.22 &  58/ 58/ 58 &  0.79 &  3.6 &   500	&   37 & 05:02:05 & 29.71/14.86/44.57 &  56/ 56/ 56 &  0.53 &  4.9 &   200 \\ 
   3 & 05:19:17 & 30.22/15.12/45.34 &  59/ 59/ 59 &  0.87 &  3.1 &   500	&   38 & 05:08:10 & 29.96/14.99/44.96 &  57/ 57/ 57 &  0.51 &  5.1 &   500 \\ 
   4 & 05:21:36 & 30.30/15.16/45.46 &  59/ 59/ 59 &  0.90 &  3.1 &   500	&   39 & 05:10:30 & 30.05/15.03/45.08 &  58/ 58/ 58 &  0.54 &  4.8 &   500 \\ 
   5 & 05:23:56 & 30.38/15.20/45.58 &  59/ 59/ 59 &  1.02 &  2.6 &   500	&   40 & 05:12:49 & 30.13/15.08/45.21 &  58/ 58/ 58 &  0.55 &  4.6 &   500 \\ 
   6 & 06:09:54 & 31.59/15.80/47.39 &  66/ 66/ 66 &  0.56 &  5.4 &   500	&   41 & 05:15:08 & 30.21/15.12/45.33 &  59/ 58/ 59 &  0.57 &  4.5 &   500 \\ 
   7 & 06:12:13 & 31.63/15.83/47.45 &  66/ 66/ 66 &  0.54 &  5.0 &   500	&   42 & 05:17:28 & 30.30/15.16/45.45 &  59/ 59/ 59 &  0.60 &  4.2 &   500 \\ 
   8 & 06:14:32 & 31.67/15.84/47.51 &  66/ 66/ 66 &  0.52 &  5.1 &   500	&   43 & 06:00:05 & 31.47/15.75/47.22 &  65/ 65/ 65 &  0.70 &  3.7 &   500 \\ 
   9 & 06:16:52 & 31.70/15.86/47.57 &  66/ 66/ 66 &  0.54 &  4.9 &   500	&   44 & 06:02:25 & 31.52/15.77/47.29 &  65/ 65/ 65 &  0.71 &  3.6 &   500 \\ 
  10 & 06:19:11 & 31.74/15.88/47.62 &  67/ 67/ 67 &  0.57 &  4.6 &   500	&   45 & 06:04:45 & 31.57/15.79/47.36 &  65/ 65/ 65 &  0.62 &  4.1 &   500 \\ 
  11 & 07:10:49 & 31.92/15.97/47.90 &  73/ 73/ 73 &  0.41 &  6.2 &   500	&   46 & 06:07:04 & 31.61/15.81/47.42 &  66/ 66/ 66 &  0.60 &  4.2 &   500 \\ 
  12 & 07:13:08 & 31.90/15.96/47.87 &  73/ 73/ 73 &  0.38 &  6.7 &   500	&   47 & 06:09:24 & 31.65/15.83/47.48 &  66/ 66/ 66 &  0.61 &  4.1 &   500 \\ 
  13 & 07:15:27 & 31.88/15.95/47.84 &  73/ 73/ 73 &  0.39 &  6.6 &   500	&   48 & 07:32:32 & 31.58/15.80/47.38 &  75/ 75/ 75 &  0.50 &  6.0 &   500 \\ 
  14 & 07:17:47 & 31.86/15.94/47.80 &  73/ 73/ 73 &  0.41 &  6.2 &   500	&   49 & 07:34:51 & 31.53/15.78/47.31 &  76/ 76/ 76 &  0.50 &  6.0 &   500 \\ 
  15 & 07:20:06 & 31.83/15.93/47.76 &  74/ 74/ 74 &  0.39 &  6.6 &   500	&   50 & 07:37:11 & 31.48/15.75/47.24 &  76/ 76/ 76 &  0.50 &  6.0 &   500 \\ 
  16 & 07:49:43 & 31.28/15.65/46.93 &  77/ 77/ 77 &  0.53 &  4.8 &   500	&   51 & 07:39:30 & 31.43/15.73/47.16 &  76/ 76/ 76 &  0.52 &  5.9 &   500 \\ 
  17 & 07:52:02 & 31.22/15.62/46.84 &  77/ 77/ 77 &  0.46 &  5.5 &   500	&   52 & 07:41:50 & 31.38/15.70/47.08 &  76/ 76/ 76 &  0.54 &  5.7 &   500 \\ 
  18 & 07:54:22 & 31.16/15.59/46.75 &  77/ 77/ 77 &  0.44 &  5.7 &   500	&   53 & 08:18:31 & 30.19/15.11/45.30 &  80/ 80/ 80 &  0.52 &  6.3 &   500 \\ 
  19 & 07:56:42 & 31.09/15.56/46.65 &  77/ 77/ 77 &  0.42 &  5.9 &   500	&   54 & 08:20:50 & 30.10/15.06/45.16 &  80/ 80/ 80 &  0.56 &  5.9 &   500 \\ 
  20 & 07:59:02 & 31.02/15.52/46.54 &  78/ 78/ 78 &  0.43 &  5.8 &   500	&   55 & 08:23:10 & 30.00/15.01/45.01 &  80/ 80/ 80 &  0.54 &  6.3 &   500 \\ 
  21 & 08:29:57 & 29.87/14.95/44.82 &  81/ 81/ 81 &  0.54 &  4.6 &   500	&   56 & 08:25:29 & 29.90/14.96/44.86 &  81/ 81/ 81 &  0.49 &  6.9 &   500 \\ 
  22 & 08:32:17 & 29.77/14.90/44.67 &  81/ 81/ 81 &  0.49 &  5.1 &   500	&   57 & 08:27:48 & 29.79/14.91/44.70 &  81/ 81/ 81 &  0.46 &  7.3 &   500 \\ 
  23 & 08:34:37 & 29.66/14.84/44.51 &  81/ 81/ 81 &  0.49 &  5.1 &   500	&   58 & 09:20:25 & ------/------/40.22 &  ---/---/ 86 &  0.67 &  5.8 &   500 \\ 
  24 & 08:36:56 & 29.55/14.79/44.34 &  81/ 81/ 81 &  0.50 &  5.0 &   500	&   59 & 09:22:44 & ------/------/39.98 &  ---/---/ 86 &  0.65 &  5.9 &   500 \\ 
  25 & 08:39:15 & 29.44/14.73/44.18 &  82/ 82/ 82 &  0.49 &  5.1 &   500	&   60 & 09:25:04 & ------/------/39.74 &  ---/---/ 86 &  0.64 &  6.1 &   500 \\ 
  26 & 09:10:08 & 27.73/13.88/41.61 &  85/ 85/ 85 &  0.40 &  6.2 &   500	&   61 & 09:27:23 & ------/------/39.49 &  ---/---/ 87 &  0.67 &  5.9 &   500 \\ 
  27 & 09:12:28 & 27.59/13.81/41.39 &  85/ 85/ 85 &  0.38 &  6.6 &   500	&   62 & 09:29:42 & ------/------/39.24 &  ---/---/ 87 &  0.71 &  5.5 &   500 \\ 
  28 & 09:14:47 & 27.44/13.73/41.17 &  85/ 85/ 85 &  0.37 &  6.7 &   500	&   63 & 10:24:35 & 21.68/10.85/32.54 &  93/ 93/ 93 &  0.56 &  8.1 &   200 \\ 
  29 & 09:17:06 & 27.29/13.66/40.95 &  85/ 85/ 85 &  0.38 &  6.6 &   500	&   64 & 10:25:37 & 21.59/10.80/32.39 &  93/ 93/ 93 &  0.59 &  7.8 &   200 \\ 
  30 & 09:19:27 & 27.14/13.58/40.72 &  85/ 85/ 85 &  0.39 &  6.4 &   500	&   65 & 10:26:39 & 21.49/10.76/32.25 &  93/ 93/ 93 &  0.60 &  7.6 &   200 \\ 
  31 & 09:48:22 & 25.05/12.54/37.59 &  88/ 88/ 88 &  0.49 &  5.0 &   500	&  &  &  &  &   &   & \\ 
  32 & 09:50:41 & 24.87/12.45/37.31 &  89/ 89/ 89 &  0.53 &  4.7 &   500	&  &  &  &  &   &   & \\ 
  33 & 09:53:00 & 24.68/12.35/37.04 &  89/ 89/ 89 &  0.55 &  4.5 &   500	&  &  &  &  &   &   & \\ 
  34 & 09:55:19 & 24.50/12.26/36.76 &  89/ 89/ 89 &  0.56 &  4.5 &   500	&  &  &  &  &   &   & \\ 
  35 & 09:57:39 & 24.31/12.17/36.48 &  89/ 89/ 89 &  0.53 &  4.7 &   500	&  &  &  &  &   &   & \\ 
\hline

\multicolumn{14}{c}{$\alpha$~Cen A} \\
\hline
\multicolumn{7}{c|}{2009 Apr 24} & \multicolumn{7}{c}{2009 Apr 25} \\
\hline
  C1 & 05:35:10 & 30.98/15.50/46.48 &  81/ 81/ 81 &  0.74 &  3.8 &   2500	&   C6 & 05:30:35 & 30.99/15.51/46.49 &  80/ 80/ 80 &  0.65 &  4.0 &   2500\\
  C2 & 06:48:04 & 29.38/14.70/44.07 &  96/ 96/ 96 &  0.77 &  3.3 &   2500	&   C7 & 06:23:17 & 29.91/14.97/44.87 &  92/ 92/ 92 &  0.60 &  4.2 &   2500\\
  C3 & 07:31:50 & 28.10/14.06/42.16 & 107/107/107 &  0.40 &  6.4 &   2500	&   C8 & 07:56:55 & 27.16/13.59/40.75 & 114/114/114 &  0.43 &  7.3 &   2500\\
  C4 & 08:10:33 & 26.84/13.43/40.27 & 116/116/116 &  0.43 &  5.8 &   2500	&   C9 & 08:41:46 & 25.66/12.84/38.50 & 125/125/125 &  0.35 &  9.7 &   2500\\
  C5 & 08:51:55 & 25.46/12.74/38.20 & 127/127/127 &  0.51 &  5.0 &   2500	&   C10 & 09:45:10 & ------/------/35.57 & ---/---/144 & 0.61 &  6.8 &  2500\\

\hline

\label{obslog2009}

\end{tabular}
\end{center}
\end{table*}

\begin{table*}
\begin{center}
\caption {Log of AMBER observations of Antares and the calibrators 
$\alpha$~Cen A and $\alpha$~Cen B in April 2010.  
The data sets \#66--99 and C\#11--C\#18 were taken with the E0-G0-H0 (16-32-48m) 
telescope configuration, while the data sets \#100--109 and C\#19 were taken 
with the E0-G0-I1 (16-57-69m) configuration.  
The seeing ($s$) and the coherence time ($\tau_0$) are in the visible.
}

\begin{tabular}{l l c c l r r | l l c c l r r}\hline
\# & $t_{\rm obs}$ & $B_{\rm p}$ & PA  & $s$ & $\tau_0$ & NDIT &
\# & $t_{\rm obs}$ & $B_{\rm p}$ & PA  & $s$  & $\tau_0$ & NDIT  \\
   &  (UTC)       &      (m)   &(\degr)&  (\arcsec)      & (ms)     &      &
   &  (UTC)       &      (m)   &(\degr)&   (\arcsec)     & (ms)     &        \\
\hline
\multicolumn{14}{c}{Antares} \\
\hline
\multicolumn{7}{c|}{2010 Apr 10} & \multicolumn{7}{c}{2010 Apr 11} \\
\hline
  66 & 06:19:19 & 45.55/15.19/30.36 &  59/ 59/ 59 &  1.02 & 47.1 &   500	&   89 & 05:47:04 & 43.95/14.66/29.30 &  55/55/55 &  0.78 & 30.5 &   500 \\ 
  67 & 06:21:30 & 45.66/15.23/30.44 &  59/ 59/ 59 &  0.97 & 31.3 &   500	&   90 & 05:49:16 & 44.09/14.70/29.39 &  55/55/55 &  0.74 & 23.9 &   500 \\ 
  68 & 06:23:41 & 45.77/15.26/30.51 &  60/ 60/ 60 &  0.99 & 28.1 &   500	&   91 & 05:51:27 & 44.22/14.75/29.47 &  55/55/55 &  0.75 & 23.1 &   500 \\ 
  69 & 06:25:51 & 45.88/15.30/30.58 &  60/ 60/ 60 &  1.11 & 25.9 &   500	&   92 & 05:53:37 & 44.35/14.79/29.56 &  56/56/56 &  0.67 & 30.7 &   500 \\ 
  70 & 06:28:02 & 45.98/15.33/30.65 &  60/ 60/ 60 &  1.07 & 29.7 &   500	&   93 & 05:55:48 & 44.48/14.83/29.65 &  56/56/56 &  0.69 & 27.5 &   500 \\ 
  71 & 07:16:22 & 47.64/15.89/31.76 &  67/ 67/ 67 &  0.83 & 21.8 &   500	&   94 & 06:34:20 & 46.44/15.49/30.95 &  62/62/62 &  0.72 & 17.1 &   500 \\ 
  72 & 07:18:32 & 47.69/15.90/31.78 &  67/ 67/ 67 &  0.81 & 21.8 &   500	&   95 & 06:36:31 & 46.53/15.52/31.01 &  62/62/62 &  0.69 & 17.8 &   500 \\ 
  73 & 07:20:43 & 47.73/15.92/31.81 &  67/ 67/ 67 &  0.78 & 20.5 &   500	&   96 & 06:38:41 & 46.61/15.54/31.07 &  62/62/62 &  0.68 & 18.2 &   500 \\ 
  74 & 07:22:53 & 47.77/15.93/31.84 &  68/ 68/ 68 &  0.84 & 17.6 &   500	&   97 & 06:40:52 & 46.70/15.57/31.13 &  63/63/63 &  0.66 & 19.5 &   500 \\ 
  75 & 07:25:03 & 47.80/15.94/31.86 &  68/ 68/ 68 &  0.85 & 18.0 &   500	&   98 & 06:43:03 & 46.78/15.60/31.18 &  63/63/63 &  0.67 & 20.3 &   500 \\ 
  76 & 08:44:42 & 46.97/15.66/31.30 &  77/ 77/ 77 &  0.68 & 15.4 &   500	&   99 & 07:20:36 & 47.79/15.94/31.86 &  68/68/68 &  0.70 & 34.6 &   500 \\ 
  77 & 08:46:53 & 46.88/15.64/31.25 &  77/ 77/ 77 &  0.66 & 15.7 &   500	&  100 & 09:21:33 & 65.11/54.08/14.96 & 118/128/81 &  0.75 & 20.6 &   500 \\ 
  78 & 08:49:03 & 46.80/15.61/31.19 &  77/ 77/ 77 &  0.69 & 15.0 &   500	&  101 & 09:23:43 & 64.94/53.97/14.91 & 119/128/81 &  0.74 & 21.7 &   500 \\ 
  79 & 08:51:14 & 46.71/15.58/31.13 &  77/ 77/ 77 &  0.71 & 14.8 &   500	&  102 & 09:25:54 & 64.77/53.85/14.86 & 119/129/81 &  0.76 & 23.1 &   500 \\ 
  80 & 08:53:25 & 46.61/15.55/31.07 &  77/ 77/ 77 &  0.69 & 14.8 &   500	&  103 & 09:28:05 & 64.60/53.74/14.81 & 119/129/81 &  0.78 & 23.4 &   500 \\ 
  81 & 09:36:33 & 44.08/14.70/29.38 &  82/ 82/ 82 &  0.70 & 15.2 &   500	&  104 & 09:30:15 & 64.42/53.62/14.76 & 120/129/81 &  0.78 & 23.4 &   500 \\ 
  82 & 09:38:45 & 43.92/14.65/29.27 &  82/ 82/ 82 &  0.85 & 12.8 &   500	&  105 & 09:56:49 & 62.04/52.07/14.04 & 124/134/84 &  0.71 & 30.4 &   500 \\ 
  83 & 09:40:55 & 43.75/14.59/29.16 &  82/ 82/ 82 &  0.88 & 12.8 &   500	&  106 & 09:58:59 & 61.83/51.93/13.98 & 124/134/84 &  0.68 & 28.1 &   500 \\ 
  84 & 09:43:06 & 43.58/14.54/29.05 &  82/ 82/ 82 &  0.85 & 12.7 &   500	&  107 & 10:01:09 & 61.61/51.79/13.91 & 125/135/84 &  0.71 & 27.2 &   500 \\ 
  85 & 09:45:17 & 43.41/14.48/28.93 &  82/ 82/ 83 &  0.91 & 10.3 &   500	&  108 & 10:03:19 & 61.40/51.66/13.84 & 125/135/85 &  0.81 & 26.8 &   500 \\ 
  86 & 10:19:23 & 40.32/13.45/26.87 &  86/ 86/ 86 &  0.64 & 27.5 &   500	&  109 & 10:05:31 & 61.18/51.52/13.77 & 126/136/85 &  0.90 & 26.3 &   500 \\ 
  87 & 10:21:34 & 40.09/13.37/26.72 &  86/ 86/ 86 &  0.68 & 26.1 &   500	&      &          &                   &             &       &       &       \\
  88 & 10:23:45 & 39.87/13.30/26.57 &  86/ 86/ 86 &  0.72 & 24.6 &   500	&      &          &                   &             &       &       &       \\
\hline

\multicolumn{14}{c}{$\alpha$~Cen A} \\
\hline
\multicolumn{7}{c|}{2010 Apr 10} & \multicolumn{7}{c}{2010 Apr 11} \\
\hline
  C11 & 05:57:07 & 47.21/15.74/31.46 &  73/ 73/ 73 &  0.85 & 35.2 &   2500	&   C16 & 05:21:35 & 47.66/15.89/31.77 &  67/ 67/ 67 & -99.99 & 24.5 &   2500 \\ 
  C12 & 06:55:08 & 45.81/15.28/30.53 &  86/ 86/ 86 &  0.78 & 30.6 &   2500	&   C17 & 06:12:28 & 46.82/15.61/31.21 &  77/ 77/ 77 &  0.67 & 20.1 &   2500 \\ 
  C13 & 07:46:14 & 43.98/14.67/29.32 &  97/ 97/ 97 &  0.67 & 21.5 &   2500	&   C18 & 07:00:07 & 45.53/15.18/30.35 &  88/ 88/ 88 &  0.58 & 31.5 &   2500 \\ 
  C14 & 09:16:12 & 39.78/13.27/26.52 & 119/118/119 &  0.71 & 13.8 &   2500	&      &          &                   &             &       &       &       \\
  C15 & 10:02:42 & 37.47/12.49/24.97 & 131/131/131 &  0.69 & 13.8 &   2500	&      &          &                   &             &       &       &       \\
\hline

\multicolumn{14}{c}{$\alpha$~Cen B} \\
\hline
\multicolumn{7}{c|}{2010 Apr 10} & \multicolumn{7}{c}{2010 Apr 11} \\
\hline
      &          &                   &             &       &       &            &   C19 & 08:58:05 & 68.07/56.52/13.50 & 143/150/115 &  0.54 & 32.9 &   2500 \\        
\hline

\label{obslog2010}
\end{tabular}
\end{center}
\end{table*}

\section{Image reconstruction of simulated data}
\label{appendix_simtests}

We generated interferometric data (visibility amplitudes and closure phases) 
from simulated stellar images by sampling at the same $u\varv$ points 
as our AMBER observations.  
The image reconstruction from these simulated interferometric data allows us 
to find the appropriate parameters (such as the initial model, prior, and 
regularization scheme) that can restore the original images correctly.  

For our data on Antares, we tested the image reconstruction for a 
limb-darkened disk with the angular diameter and the limb-darkening parameter 
derived for the 2009 data 
and a uniform disk with surface inhomogeneities and an 
extended component.  
The simulated interferometric data were generated using 
a program developed by one of the authors (K.-H. Hofmann) as follows.  
For each telescope triplet, we simulated two-telescope interferograms 
corresponding to the three baseline vectors for a given object 
(e.g., limb-darkened stellar disk or 
spotted star).  
The simulated interferograms were degraded by the atmospheric piston, 
photon noise, 
sky background, and detector noise.  From $\sim$1000 simulated two-telescope 
interferograms on each baseline, 
the average power spectrum and bispectrum were calculated. 
The subtraction of the noise bias terms and calibration with an unresolved 
calibrator star yielded the calibrated visibilities and closure phases 
for the simulated object.  
The amount of noise in the simulated interferograms was chosen to obtain 
approximately the same errors as in the AMBER measurements of Antares.

The reconstructed images were convolved with the clean beam, which was 
determined in the following manner.  The spatial resolution in the direction 
of the $u\varv$ points with the longest baseline length ($B_{\rm max}$) is given 
by $\lambda/B_{\rm max}$ = 9.8~mas.  However, the sparse $u\varv$ coverage shown 
in Fig.~\ref{uv_coverage} leads to a strongly elongated beam.  We fitted the 
central peak of the dirty beam with a 2-D Gaussian and derived the 
ratio between the major and minor axes and the position angle.  
The beam defined in this manner is narrower than that given by 
$\lambda/B_{\rm max}$, because of the lack of data at very short baselines.  
Therefore, we scaled the major and minor axes of this Gaussian beam so that 
the beam size along the minor axis matches the resolution given by 
$\lambda/B_{\rm max}$.  The final clean beam for the $u\varv$ coverage 
obtained in 2009 is a 2-D Gaussian with 
$9.8\times63.2$~mas (FWHM) with the major axis at a position angle of $-12$\degr.

\begin{figure*}
\resizebox{\hsize}{!}{\rotatebox{-90}{\includegraphics{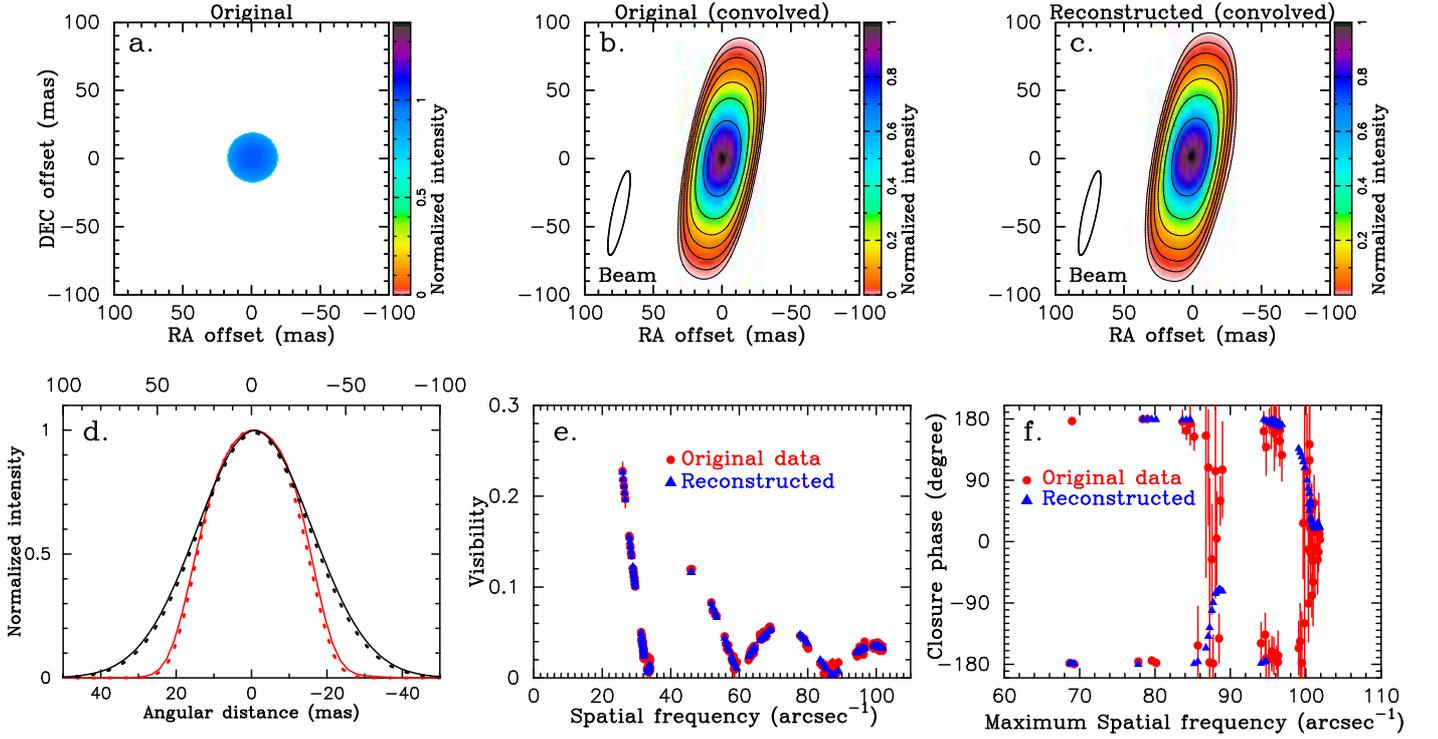}}}
\caption{
Image reconstruction from the simulated data for a limb-darkened disk with the 
parameters derived from the 2009 data.  
{\bf a:} Original image. 
{\bf b:} Original image after convolving with the Gaussian beam with  
$9.8 \times 63.2$~mas.  
{\bf c:} Reconstructed image after convolving with the beam. 
{\bf d:} The intensity profiles of the original image are plotted by the red 
dotted line (along the minor axis) and the black dotted line (along the 
major axis).  
The intensity profiles of the reconstructed images are plotted by the red 
solid line (along the minor axis) and black solid line (along the major axis).  
The abscissa for the intensity profiles along the major axis is shown at the 
top of the panel. 
{\bf e:} Comparison between the visibilities generated from the original image 
(filled dots) and those from the reconstructed image (triangles). 
{\bf f:} Comparison between the CPs generated from the original image 
(filled dots) and those from the reconstructed image (triangles). 
}
\label{simtest_ldd}
\end{figure*}

\begin{figure*}
\resizebox{\hsize}{!}{\rotatebox{-90}{\includegraphics{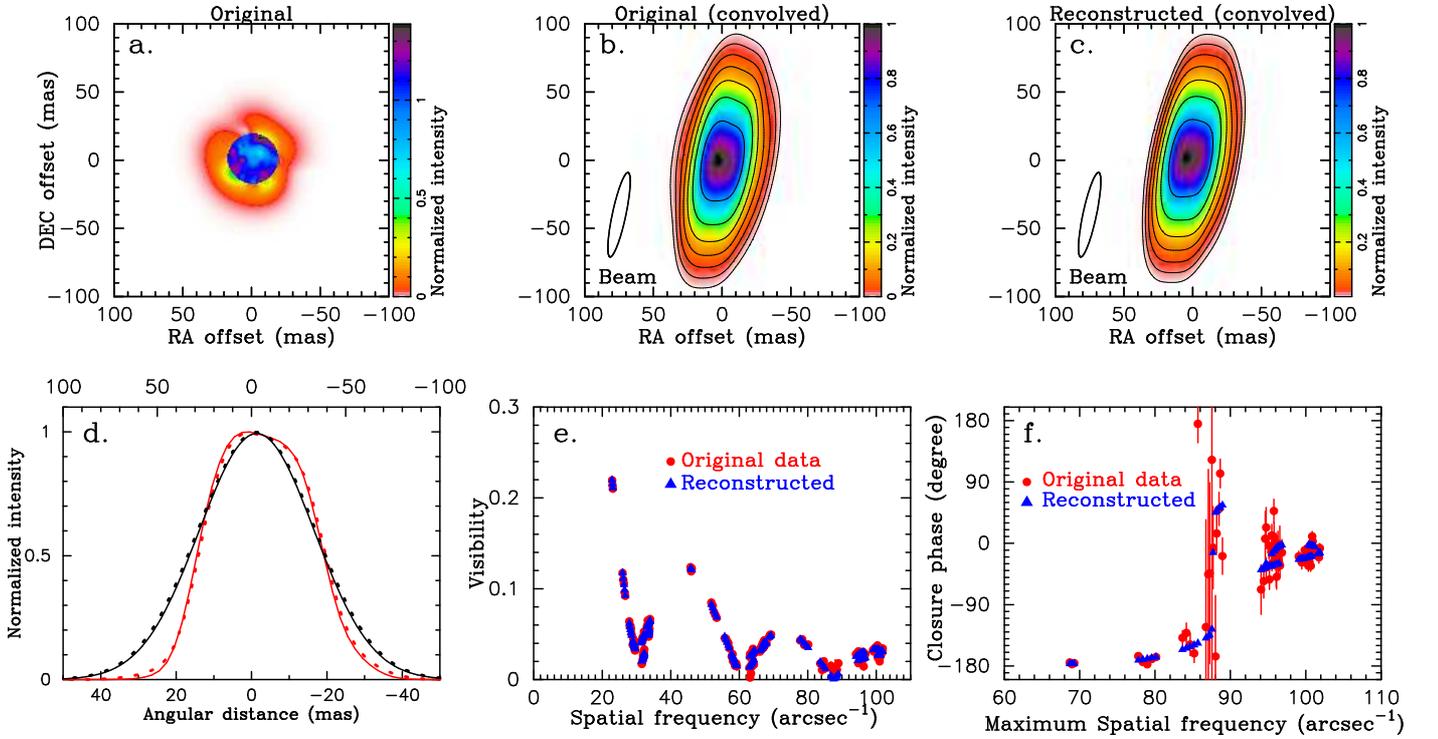}}}
\caption{
Image reconstruction from the simulated data for a uniform disk with surface 
inhomogeneities and an asymmetric extended component, shown in the same manner 
as Fig.~\ref{simtest_ldd}
}
\label{simtest_spot}
\end{figure*}

The results of the image reconstruction of two sets of simulated data are 
shown in Figs.~\ref{simtest_ldd} and \ref{simtest_spot}.  The reconstructed 
images agree reasonably well with the original image after convolving with 
the beam.  
These tests show that the uniform-disk initial model with angular diameters 
of 36--37~mas and the same prior as used in Paper II are appropriate for the 
image reconstruction from our AMBER data of Antares.  
The prior is given by 
\[
Pr(r) = \frac{1}{e^{(r-r_{\rm p})/\varepsilon_{\rm p}} + 1},
\]
where $r$ is the radial coordinate in mas, and $r_{\rm p}$ and $\varepsilon_{\rm p}$ 
define the size and the smoothness of the edge, respectively 
($\varepsilon_{\rm p} \rightarrow 0$ corresponds to a uniform disk).  
The appropriate values for $r_{\rm p}$ and $\varepsilon_{\rm p}$ were found to
be 12~mas and 2.4~mas, respectively.  We adopted the maximum entropy 
regularization as in Paper II.  

\section{Fit to the interferometric data}
\label{appendix_fit}

\begin{figure*}
\resizebox{\hsize}{!}{\rotatebox{-90}{\includegraphics{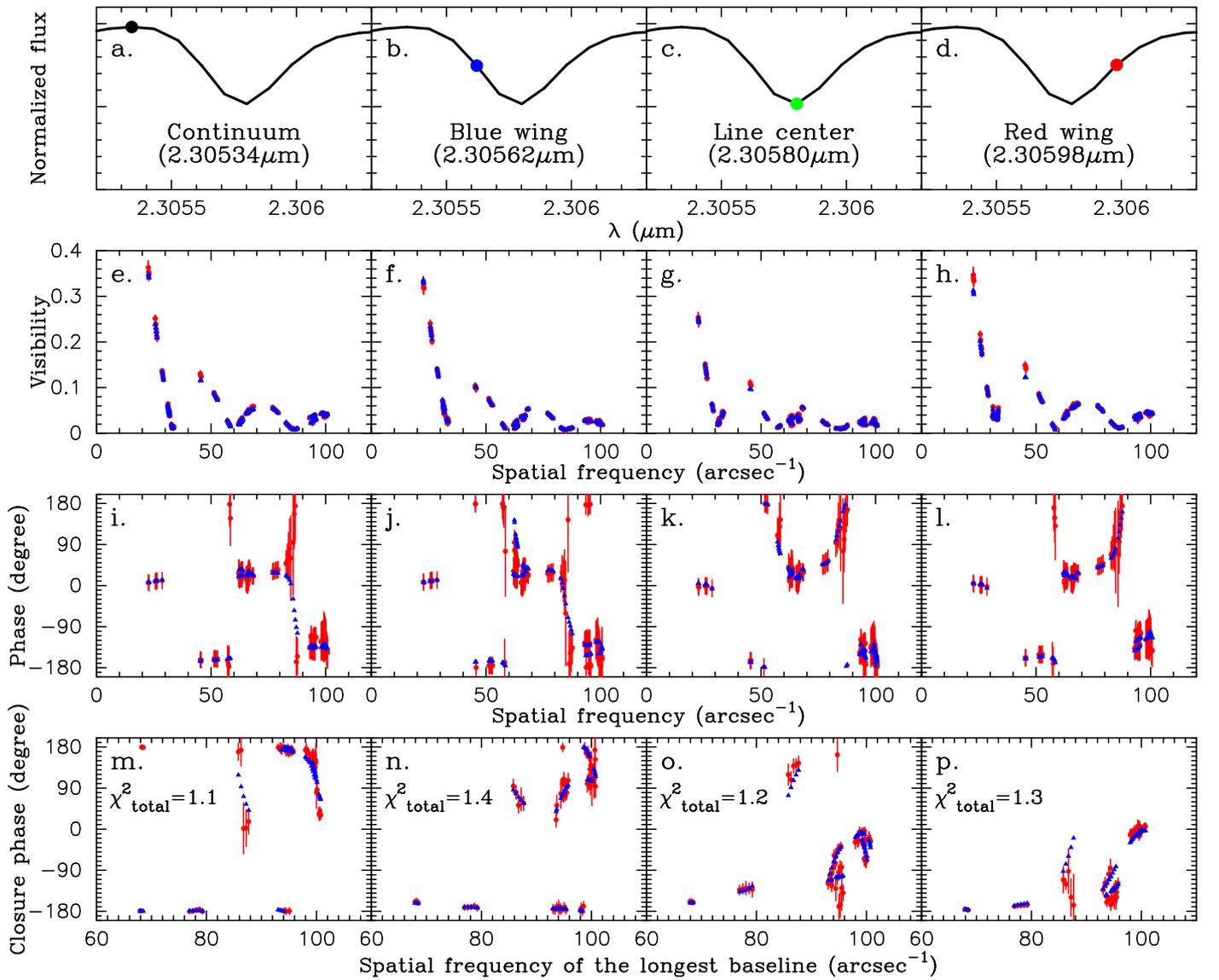}}}
\caption{
Comparison between the interferometric observables measured in 2009 and those 
from the reconstructed images within the CO line profile shown in 
Fig.~\ref{images2009}.  The first, second, third, and fourth columns show the 
comparison for the continuum, blue wing, line center, and red wing, 
respectively.  The filled circles in the top row ({\bf a}--{\bf d}) show the 
positions within the CO line profile.  In the remaining panels, the observed 
data and those from the image reconstruction are plotted by the dots and 
triangles, respectively.  The reduced $\chi^2$ values including the 
visibilities, phases, and closure phases, are given in the panels in the 
bottom row.  
}
\label{fit_data2009}
\end{figure*}

\begin{figure*}
\resizebox{\hsize}{!}{\rotatebox{-90}{\includegraphics{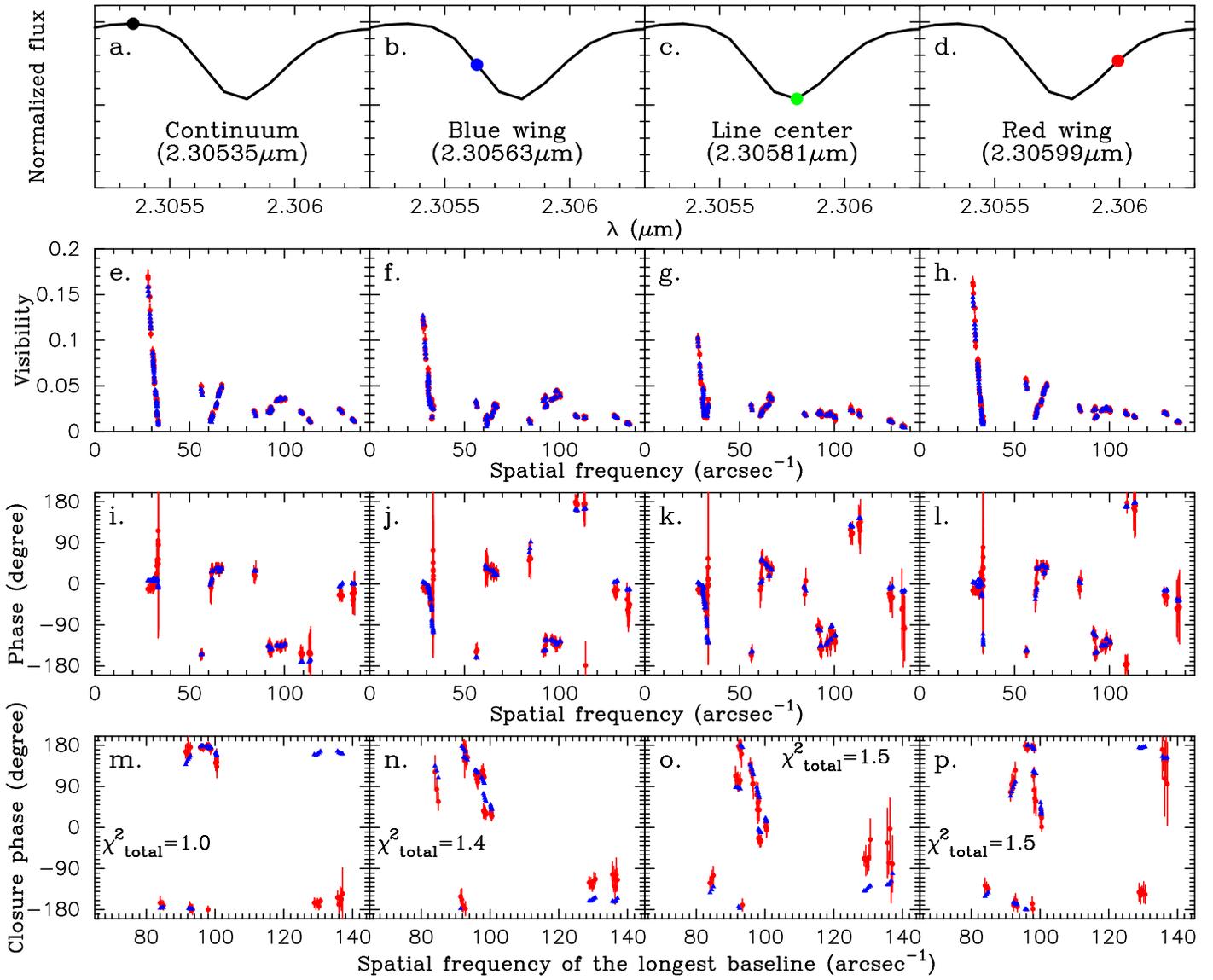}}}
\caption{
Comparison between the interferometric observables measured in 2010 
and those from the reconstructed images within the CO line profile shown in 
Fig.~\ref{images2010}.  The panels are shown in the same manner as in 
Fig.~\ref{fit_data2009}. 
}
\label{fit_data2010}
\end{figure*}

Figures~\ref{fit_data2009} and \ref{fit_data2010} show the fit to the measured 
interferometric observables for the image reconstruction in the CO line shown 
in Figs.~\ref{images2009} and \ref{images2010}, respectively.

\end{document}